\pdfoutput=1
\documentclass[journal]{vgtc}                %

\usepackage{mathptmx}
\usepackage{graphicx} %
\usepackage{times}

\usepackage[usenames,dvipsnames]{xcolor}
\usepackage{soul}

\usepackage[colorinlistoftodos,disable]{todonotes}
\usepackage{ifthen}
\usepackage{setspace}

\reversemarginpar %

\newcommand{\remark}[4]{%
\ifthenelse{\equal{#4}{m}}%
{%
\todo[caption=xxx,color=#3,size=\small]{%
\begin{spacing}{0.9}%
\textsf{\textbf{#1:}\\#2}%
\end{spacing}%
}%
}{
\begin{spacing}{1.2}
\todo[inline, color=#3,size=\small,caption=xxx]{
\textsf{\textbf{#1:}~#2}
}
\end{spacing}
}
}

\definecolor{remarkSarah}{HTML}{FED9A6}
\definecolor{remarkTim}{HTML}{B3CDE3}
\definecolor{remarkKim}{HTML}{CCEBC5}
\definecolor{remarkYalong}{HTML}{DECBE4}
\definecolor{remarkReviewer}{HTML}{FBB4AE}

\definecolor{remarkLang}{HTML}{FB9A99}
\definecolor{remarkLangQ}{HTML}{FDB462}

\newcommand{\yy}[2][inline]{\remark{YY}{#2}{remarkYalong}{#1}}
\newcommand{\rev}[2][inline]{\remark{Reviewer}{#2}{remarkReviewer}{#1}}

\usepackage{booktabs}
\usepackage[none]{hyphenat}
\usepackage{subfig}

\usepackage{multirow}

\usepackage{tabularx}
\newcommand{\tabincell}[2]{\begin{tabular}{@{}#1@{}}#2\end{tabular}}

\usepackage{array}
\newcolumntype{L}[1]{>{\raggedright\let\newline\\\arraybackslash\hspace{0pt}}m{#1}}
\newcolumntype{C}[1]{>{\centering\let\newline\\\arraybackslash\hspace{0pt}}m{#1}}
\newcolumntype{R}[1]{>{\raggedleft\let\newline\\\arraybackslash\hspace{0pt}}m{#1}}

\usepackage{enumitem}
\setlist{nolistsep}

\usepackage[bookmarks,backref=true,linkcolor=black]{hyperref} %
\hypersetup{
  pdfauthor = {},
  pdftitle = {},
  pdfsubject = {},
  pdfkeywords = {},
  colorlinks=true,
  linkcolor= black,
  citecolor= black,
  pageanchor=true,
  urlcolor = black,
  plainpages = false,
  linktocpage
}

\onlineid{0}

\vgtccategory{Research}

\vgtcinsertpkg

\title{Many-to-Many Geographically-Embedded\\ Flow Visualisation: An Evaluation}
\author{Yalong Yang, Tim Dwyer, Sarah Goodwin and Kim Marriott}

\authorfooter{
Yalong Yang and Kim Marriott are with Monash University and Data61, CSIRO, Victoria. E-mail: \{yalong.yang, kim.marriott\}@monash.edu.

Tim Dwyer and Sarah Goodwin are with Monash University, E-mail: \{tim.dwyer, sarah.goodwin\} @ monash.edu.

}

\shortauthortitle{Yang \MakeLowercase{\textit{et al.}}: Many-to-Many Geographically-Embedded Flow Visualisation}

\abstract{
Showing flows of people and resources between multiple geographic locations is a challenging visualisation problem.  We conducted two quantitative user studies to evaluate different visual representations for such dense many-to-many flows. In our first study we compared a bundled node-link flow map representation  and OD Maps~\cite{Wood:2010be} with a new visualisation we call MapTrix. Like OD Maps, MapTrix  overcomes the clutter associated with a traditional flow map while providing geographic embedding that is missing in standard OD matrix representations. We found that OD Maps and MapTrix had similar performance while bundled node-link flow map representations did not scale at all well. Our second study compared participant performance with OD Maps and MapTrix on larger data sets. Again performance was remarkably similar.
} %


\CCScatlist{ %
 \CCScat{H.5.2}{Information Interfaces and Presentation
  (e.g. HCI)}{User Interfaces}
}

  \teaser{
      \centering\subfloat[Bundled Flow Map]{\label{subfig:arrow-au}\includegraphics[width=0.30\textwidth]{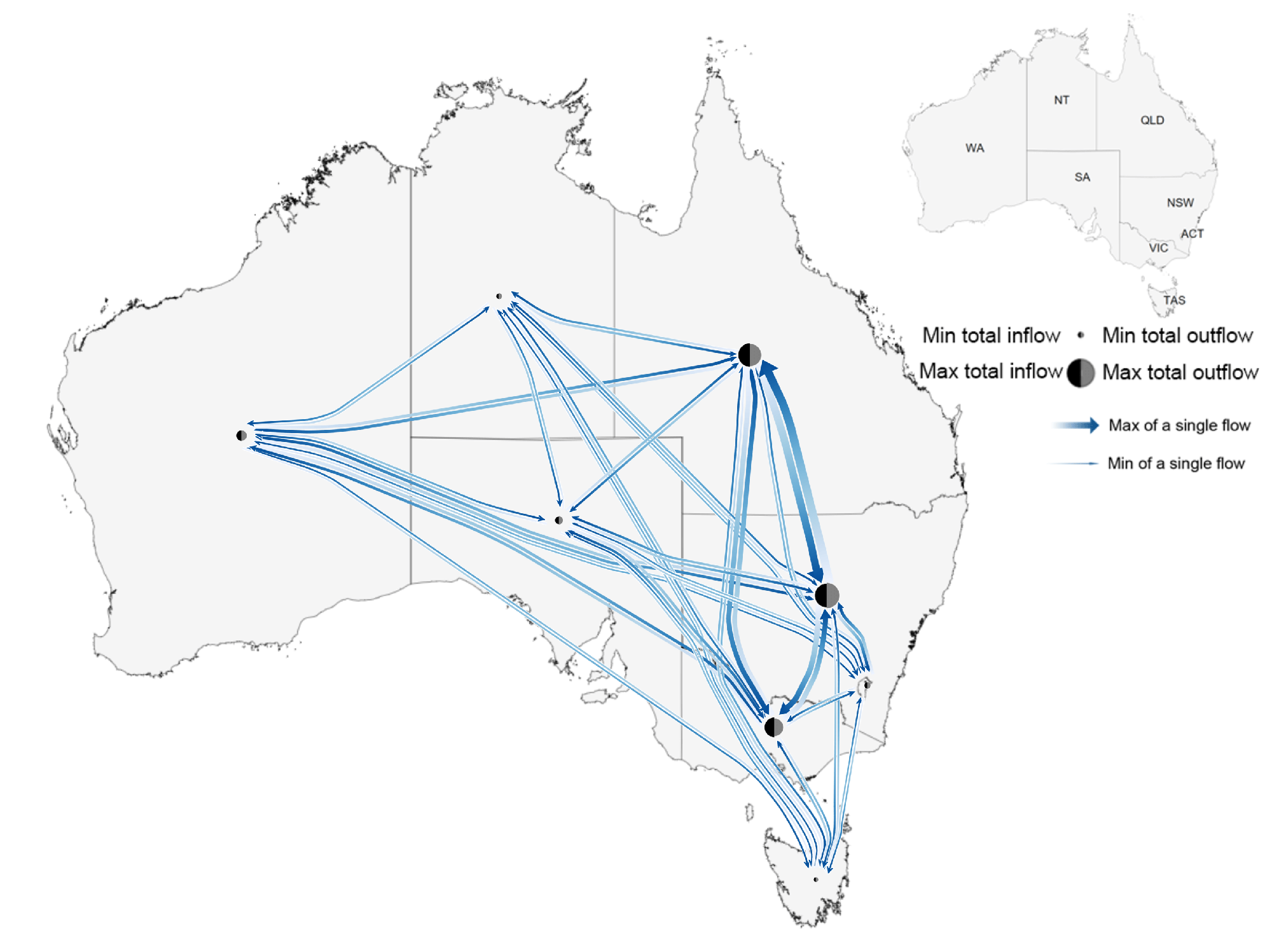}}
      \centering\subfloat[OD Map]{\label{subfig:grid-au}\includegraphics[width=0.33\textwidth]{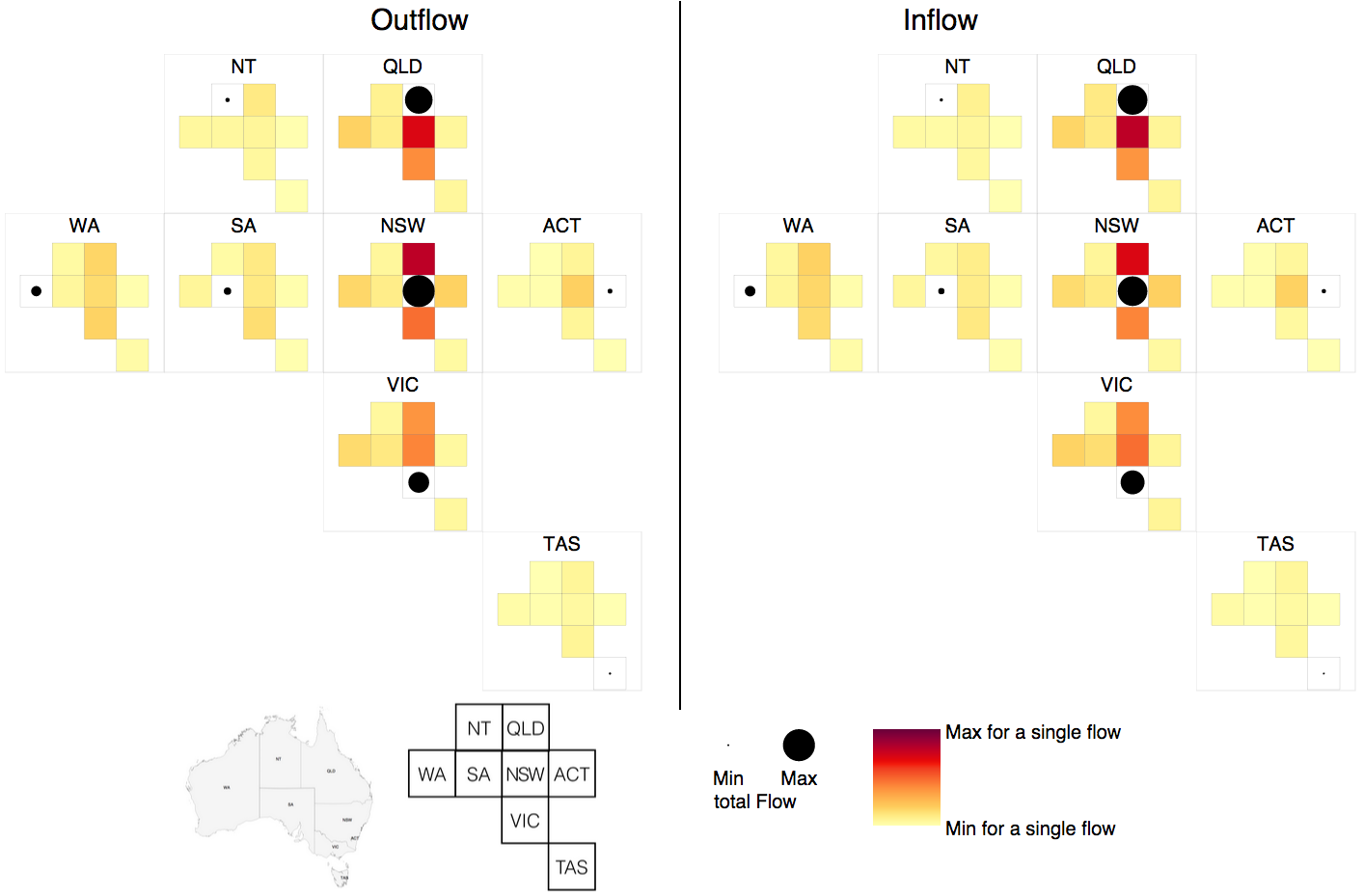}}
      \centering\subfloat[MapTrix]{\label{subfig:maptrix-au}\includegraphics[width=0.36\textwidth]{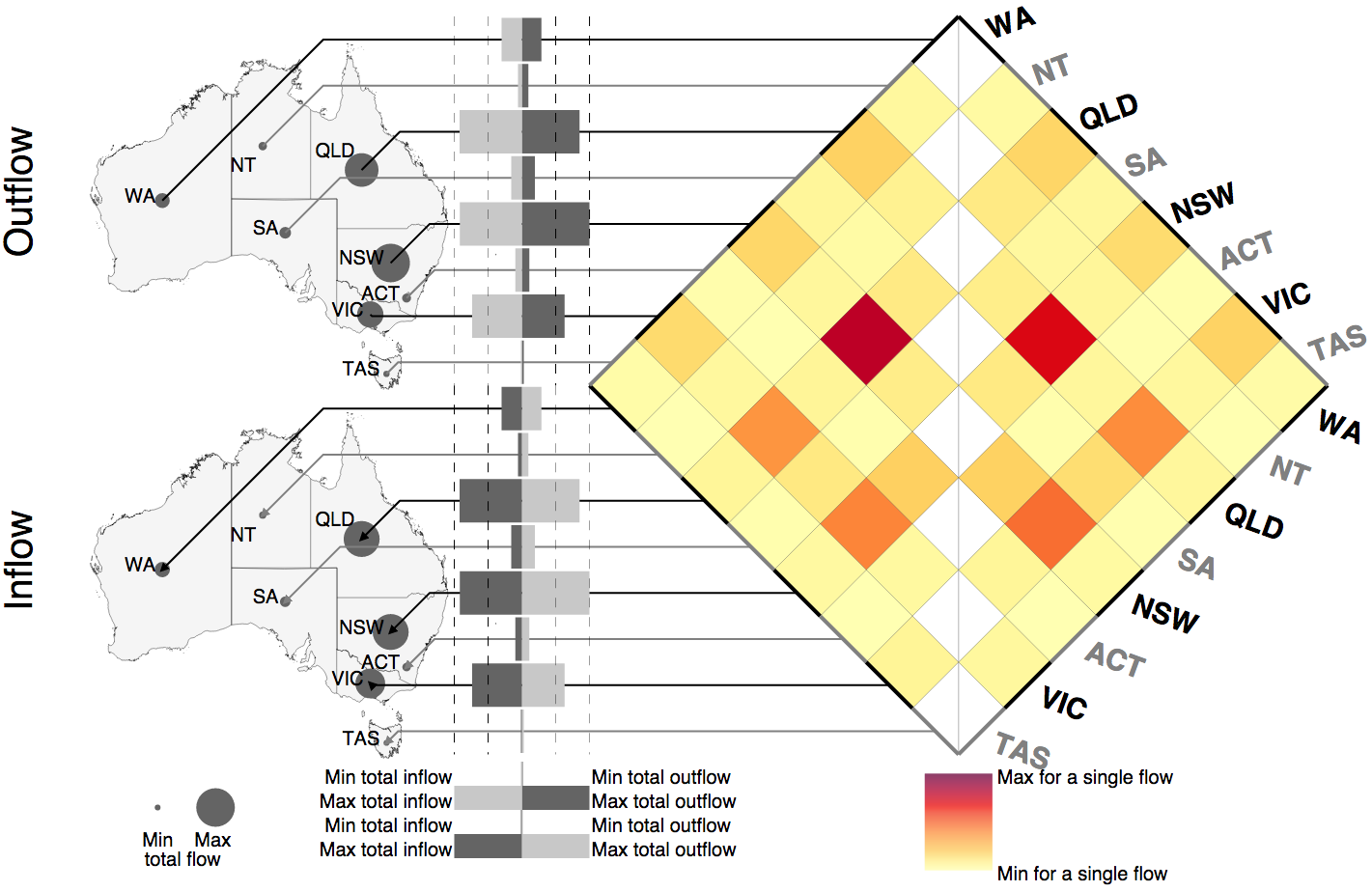}}
      \caption{The three visualisation methods compared in our first user study.  Australia was 1 of 3 countries tested. The MapTrix visualisation shown on the right and the two user studies are the main contributions of this paper. 
      \label{fig:threevis-au}}
  }

\begin{document}
\firstsection{Introduction}
\maketitle

In many applications it is important to visualise the flow of some kind of commodity between different geographic locations. Such many-to-many flows include---for example---movement of animals or disease~\cite{Gilbert:2005bd, Guo:2007gi},  movement of goods or knowledge~\cite{Paci:2008iq}, migration patterns~\cite{Tobler:1987kn} and commuting behaviour~\cite{Chiricota:2008ut}. There are two main approaches to visually presenting such flows: \emph{flow maps} and \emph{matrices}. 

Flow maps present origins and destinations on a map connected by lines or arrows. Whilst flow maps are intuitive and are well suited to show the flows from a single source, they quickly become cluttered and difficult to read when the number of commodity sources increases. 

For the matrix approach, the most basic is the origin-destination (OD) matrix~\cite{Voorhees:2013cx} in which there is a row $r$ for each source (origin), a column $c$ for each destination, and a cell $(r,c)$ shows the flow from source to destination. The drawback of the OD matrix is that geographical embedding of the sources and destinations is missing. This can be partly ameliorated by ordering the rows and columns by location, say east to west. More recently \emph{OD maps}~\cite{Wood:2010be} have been suggested. These  use a gridded two-level spatial treemap to cleverly preserve the approximate spatial (map) layout~\cite{Wood:2008}, e,g. see Fig.~\ref{fig:threevis-au}(b).

Given the practical importance of understanding many-to-many commodity flows, it is surprising that---to the best of our knowledge---there have been no user studies comparing different visualisations for showing flow. The most relevant study is by Ghoniem \textit{et al.}~\cite{Ghoniem:2004jk} which demonstrates that a  matrix representation of a general network performs better than a node-link diagram for large or dense datasets. However, their evaluation does not consider the specific application of commodity flows and the need to embed this in a geographic context.

This paper's main contributions are two-fold: (1) a new hybrid visualisation method, \emph{MapTrix}, for showing flow that combines the OD matrix and flow map representations, preserving the benefits of both;
and (2), to the best of our knowledge, the first quantitative user studies to compare different (static) visual representations of dense many-to-many commodity flows.

An example of our novel MapTrix design is shown in Fig.~\ref{fig:threevis-au}(c). In MapTrix the OD matrix is embedded spatially by using leader lines to link each row and column with its geographic location on a map. The leader lines remove the clutter of standard flow maps while still showing the position of the sources and destinations.  
We present our design decisions, alternative  variations and the algorithm for computing the MapTrix representation, where the primary issue is ensuring that the leader lines do not overlap. Our algorithm uses the one-sided boundary labelling model of Bekos \emph{et al.}~\cite{Bekos:2007hn}. However this can lead to overlapping and difficult-to-read leader lines: in a second step we increase the separation between leader lines by solving a novel quadratic program to adjust their position.

As discussed in the next section, flow maps using arrows become very cluttered when depicting a large number of flows between multiple sources and destinations.  In order to study denser flow maps we adapt a state-of-the-art ``bundling'' technique from the field of network visualisation, as described in Section~\ref{sec-bfmd}.

We conducted two  quantitative user studies. The first investigated user preferences and task performance for three very different visualisations: bundled flow map, OD map and MapTrix. Example stimuli from the study are shown in Fig.~\ref{fig:threevis-au}. 62 participants completed our on-line questionnaire. We found that MapTrix was the preferred representation while MapTrix and OD maps had very similar task performance which was much better than  with the bundled flow map. 

In our second user study we compared MapTrix and OD maps on larger dense data sets with up to 51 sources and destinations.  At this scale it was infeasible to use the bundled flow map representation.  49 participants completed this on-line questionaire. Again, we found that task performance was very similar.  For both representations it was very difficult to compare aggregated flows between or within regions comprising several sources or destinations. 

The results of our studies provide strong guidance on how interaction could be used to improve task performance with the different representations. We discuss this more fully in Section~\ref{sec-interaction}.

\section{Related Work}
\label{sec:related-work}
\yy{
	In \textbf{summary review}; R1: an early study about flow maps; \cite{Johnson:1998hb}: This study compared a paper map series, a computer map series, and animated maps of the same data to assess the effectiveness of each technique for memorizing data symbolized by graduated flow lines. Subjects were asked to study the maps and to memorize two types of information:quantity data at specified locations on the maps and trend patterns that occurred over the maps. Analysis of response times and accuracy rates for these questions suggest that animation does not improve learning ability for quantity evaluations. It does appear, however, to improve subjects' abilities to learn and remember trend patterns in the data. Results also indicate gender differences in using animated maps. Females preferred the paper map series and completed tasks signifi- cantly more accurately with them, while males appeared to learn better with animation. Average reaction times for males were significantly faster with animation. Accuracy rates, however, failed to show a significant increase over the paper map series.
}

\yy{
	In \textbf{summary review}; R4: An edge bundling algorithm that drawing the lines separately; \cite{Nocaj:2013gd}: We therefore propose methods that bundle edges at their ends rather than their interior. This way, tangents at vertices point in the general direction of all neighbors of edges in the bundle, and ambiguity is avoided altogether. For undirected graphs our approach yields curves with no more than one turning point. For directed graphs we introduce a new drawing style, confluent spiral drawings, in which the direction of edges can be inferred from monotonically increasing curvature along each spiral segment.
}

The presentation of multiple flows on a map is a classic problem in cartography and geographical visualisation. 
Here, we discuss the three broad visualisation approaches outlined in our introduction.

\vspace*{1mm}
\noindent\textbf{Flow Map Approaches.}
The earliest known flow map was created by Henry Drury Harness in 1837 to show rail usage~\cite{Robinson:1955hz}. Shortly after, Charles Joseph Minard popularised their use with sophisticated depictions of emigration and trade~\cite{Robinson:1967cj}. In 1981 Tobler produced and tested the first computer generated node-link flow maps~\cite{Tobler:1981kw}.
Each flow was presented as a straight-line arrow connecting its origin and destination, with arrow thickness proportional to its quantity. 
Unfortunately visual clutter and line crossings are inevitable even in small datasets.
The term ``flow map'' has also been used in a very literal sense to depict flows in rivers (with a single source and single destination) on geographic maps \cite{Johnson:1998hb}.

Design strategies to alleviate clutter on flow maps have long been discussed in popular visual design and cartography texts, e.g.~\cite{bertin:1967,Dent:2008vd}.  Rae~\cite{Rae:2009iq} tests the limits of scalability of traditional flow maps with straight-line arrows, trying to make aggregate flow information visible through overlaid density maps.  Whilst these show overall trends, aggregation
and vector fields lose potentially important information about individual flows. 

Another approach is to bundle links together. Several elegant and
sophisticated `bundling' strategies have also been proposed for flows from a
single source where a simple hierarchy is possible~\cite{Phan:2005cn} and algorithms have been
developed for single-source flows that achieve aesthetic branching properties~\cite{Nocaj:2013gd, Buchin:2011fk}. While these bundling strategies are well suited for maps presenting flows for one-to-many locations, their application for many-to-many flow is limited.  To create readable node-link flow maps for
our study we adapted a bundling method originally intended for network
visualisation~\cite{pupyrev2012edge} that is capable of handling many-to-many flows, see Section \ref{sec-bfmd}.

\rev{
	R1: If the authors were able to point to any past research that demonstrates a finding about non-interactive visualization to still be relevant once interaction is included, that would strengthen their argument for doing the constrained study.
}
Interaction is another way to overcome such visual clutter.  A recent system described by van den Elzen and van Wijk \cite{van2014multivariate} provided  interactive filtering and aggregation to interactively restrict the set of origins and destinations to something manageable with an otherwise conventional flow-map representation.  Obviously, in printed or public displays such interaction is unavailable, yet
even with interaction each individual view should ideally be as informative and unambiguous as possible with respect to the underlying data~\cite{nguyen2013faithfulness}.   Thus, our primary focus in this paper is on the design of flow representations that are as readable as possible from a single view.  However, even the best possible design has limits to its scalability and so we consider interaction for novel flow representations in Section \ref{sec-interaction}.

\begin{figure}
\setlength{\abovecaptionskip}{.1cm}
\setlength{\belowcaptionskip}{-0.5cm}
\centering
    \includegraphics[width=0.86\columnwidth]{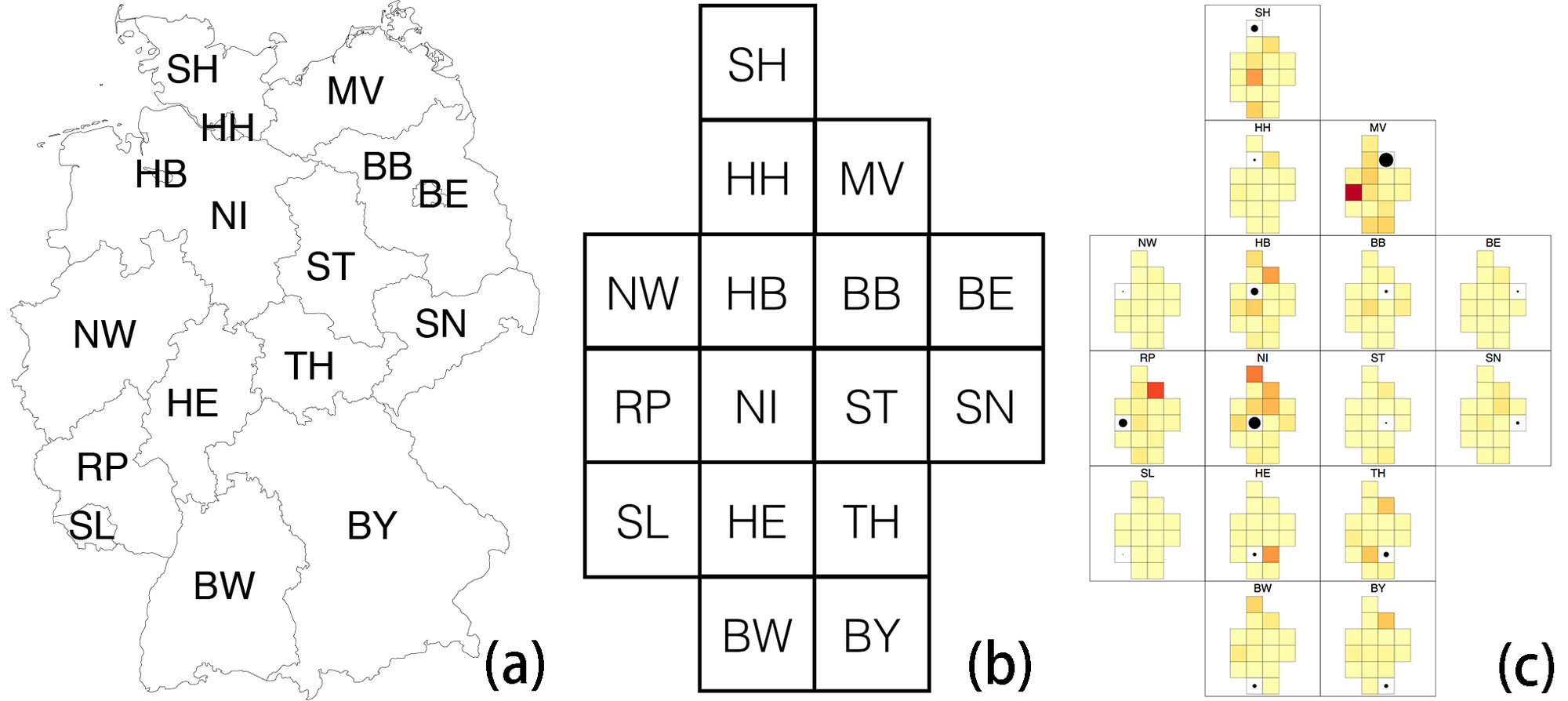}
    \caption{Demonstrating the OD map design for Germany showing abbreviations for all 16 states (a) standard map with administrative boundaries; (b) empty OD map layout; and (c) nested OD map coloured by example flow data.}
        \label{fig:ODmapLayout} 
\end{figure}

\vspace*{1mm}
\noindent\textbf{OD Matrix Based Approaches.}
Adjacency matrix representation of flow networks are called OD matrices. These present flow using a table where rows and columns represent origin and destination locations and each cell indicates the quantity of movement from one location to another. The original OD matrix dates from 1955~\cite{Voorhees:2013cx}. 
More generally, adjacency matrices have long been useful for presenting a network of relationships in a compact and structured format where reordering of rows or columns can reveal patterns~\cite{bertin:1967}.  A user study by Ghoniem~\cite{Ghoniem:2004jk} found that adjacency matrices perform better than node-link diagrams for quickly reading adjacencies. 
Whilst the original OD matrices were purely numerical, colour shading using a heatmap approach~\cite{Wilkinson:2009et} is often used to encode the size of the flow. 

One major drawback of the classic OD matrix is that it lacks a mapping from OD locations to  geographical positions. The identification of geographically related rows, columns or cells can be difficult and so spatial patterns in the dataset can be hard to determine~\cite{Wood:2010be}. 
Marble \emph{et al.}~\cite{marble1997recent} attempt to preserve the spatial properties of the OD locations by reordering columns and rows by approximate spatial position but only
limited spatial information is retained due to dimension reduction.

\emph{OD maps}~\cite{Wood:2010be} attempt to overcome this limitation through a nested small-multiples design, see Fig.~\ref{fig:ODmapLayout}.  They provide schematic geographical information by dividing the canvas into a regular grid based on the actual geographical locations on the map using a spatial treemap structure~\cite{Wood:2008}.  A second level of spatial treemaps is embedded within the first to present the OD information using colour shading. To aid readability, some cells of the grid may be left blank to indicate the outline shape of the country, e.g. Fig.~\ref{fig:ODmapLayout}(b). 
Like the OD matrix, spatial locations in OD Maps are presented as squares. As all locations have similarly sized cells, this allows data for small, highly populated areas to be seen at the same level of detail as more sparsely populated larger regions, e.g.\ in Fig.~\ref{fig:ODmapLayout} compare Berlin (BE) to Brandenburg (BB). 

Whilst spatial treemaps are currently being tested for their performance in a number of tasks~\cite{Ali:2013tg}, OD maps have yet to be evaluated in a quantitative user-study. Less formal studies show they are useful for presenting geographical commodity flows to data experts~\cite{kelly:2013, wood:2011}, but OD maps have not yet been tested on a wider audience or compared with other visualisations.
We would expect OD maps to best suit countries with similar width and height such as Ireland, Germany or Australia, while they may be less suitable for countries with elongated proportions such as Japan or New Zealand where the distortion of map location to grid location may cause cognitive difficulties.

\vspace*{1mm}
\noindent\textbf{Other Approaches.}
There have been a number of recent examples which combine other visualisations with maps to present  flow data. \emph{VIS-STAMP}~\cite{Guo:2006}, for instance, presents a matrix of small multiple maps in their  approximate spatial location with  linked views including parallel plots.

\emph{Flowstrates} connects a temporal heatmap with two maps presenting the geographical locations of origin and destination~\cite{Boyandin:2011fa} and shows how flow changes over time.  The resulting visual representation is superficially  similar to the MapTrix visualisation presented in Section~\ref{sec:maptrix}. However, while Flowstrates does present OD data it is not designed to present a complete OD matrix as we do in MapTrix. In Flowstrates each row corresponds to a single flow as it uses columns for the temporal scale. In contrast, a single MapTrix cell corresponds to a single flow. Thus to show all flows between M sources and N destinations Flowstrates requires $M\times N$ leader lines but MapTrix only $M+N$ leader lines. Furthermore, it is possible to avoid leader line crossings with MapTrix but not with Flowstrates for larger multi-way flows.

\section{Design and Implementation of MapTrix}
\label{sec:maptrix}

Our novel flow visualisation, \emph{MapTrix}, is intended to show quantitative multi-source flow data together with its associated geographical information. It has three main components: an origin map, a destination map, and an OD matrix with a single line connecting each origin and destination to the corresponding matrix row or column.

\subsection{Design of the Visual Representation}

Our first attempt to connect the OD matrix to the two maps ordered rows and columns by their map locations' $y$- and $x$-coordinate, respectively and used straight line \emph{leaders} connected map locations to their corresponding matrix row or column.  Unfortunately, this resulted in many leader line crossings making it very difficult to link rows and columns with locations.

\begin{figure}
\vspace{-8pt}
\setlength{\abovecaptionskip}{.1cm}
\setlength{\belowcaptionskip}{-0.4cm}
\includegraphics[height=38mm]{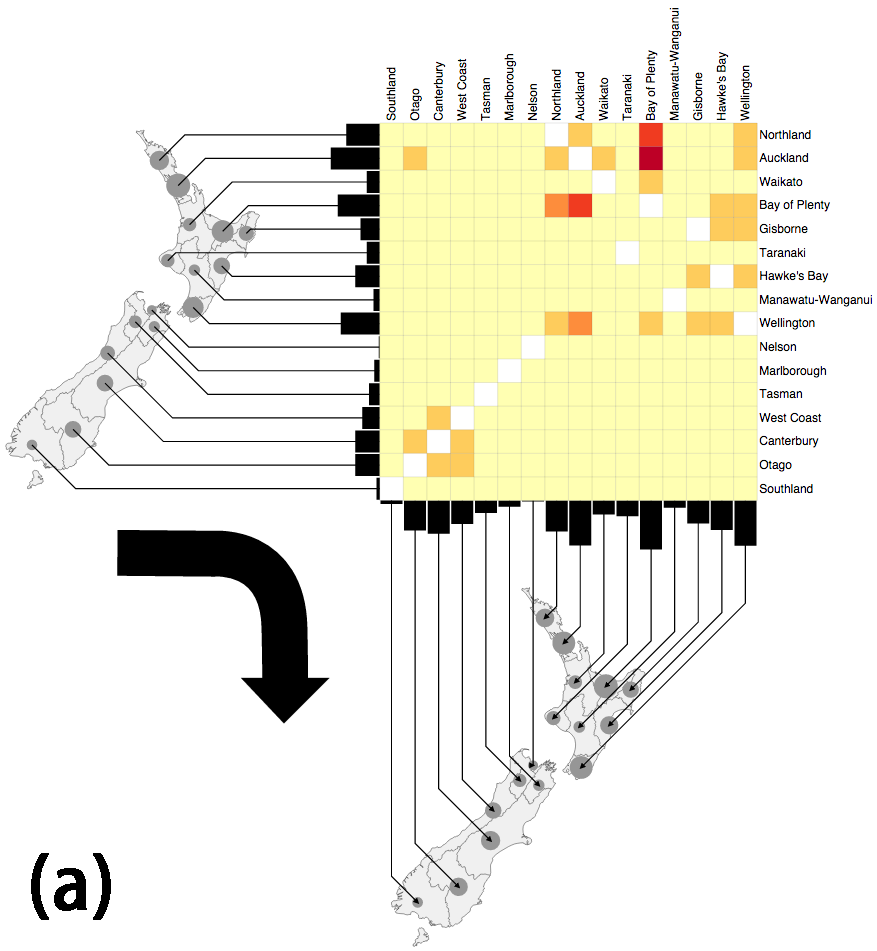}
\includegraphics[height=38mm]{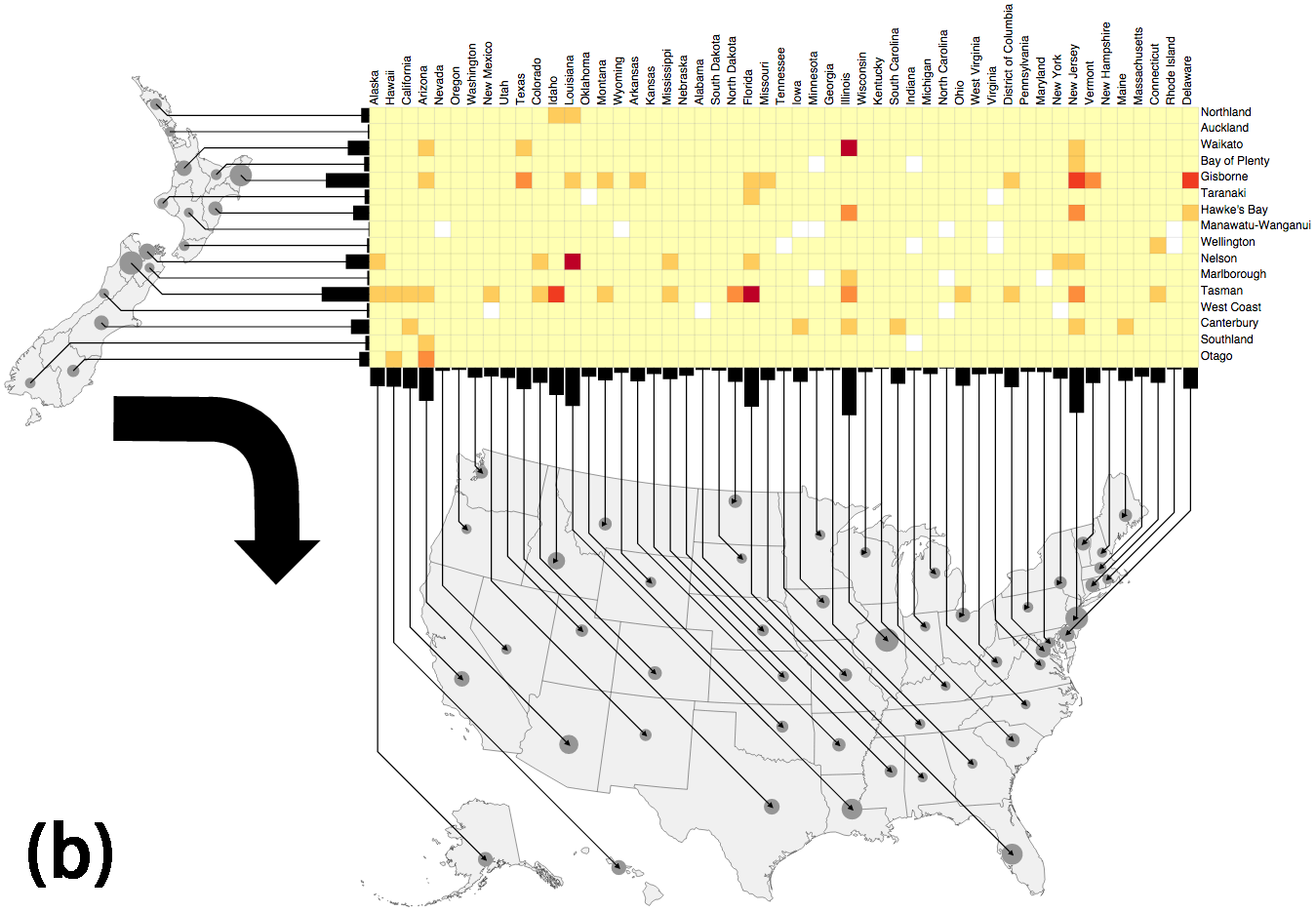}
    \caption{Intermediate designs of crossing-free leader lines connecting maps and matrix with non-matching column and row order; (a) OD flows within the same country -- New Zealand (NZ); and (b) OD flows between two different countries (NZ to USA).}
    \label{fig:maptrix-no-crossing}
\end{figure}

Our second attempt led to the design shown in Fig.~\ref{fig:maptrix-no-crossing}(a) which ensures that the connection between maps and matrix was \textit{clear, easy to track and unambiguous}. 
To achieve this we solved a so-called \emph{boundary labelling} problem which finds an ordering for the matrix rows and columns that permits leaders to connect map locations without crossings. There are various models for aesthetic boundary labelling for different situations \cite{Bekos:2009id,Bekos:2007hn,Bekos:2010tp}. Our design uses a \emph{one-sided boundary labelling model} to generate crossing-free connections with a horizontal and a diagonal segment between points in the figure and labels at one side of the figure. We introduce a novel leader adjustment algorithm (next section) to more evenly space the leader lines.

Our design uses colour shading (``YlOrRd'' continuous scale from colorbrewer~\cite{Harrower:2003jm}) to show magnitude of flow between states. Geographical locations' total in/out flows are indicated by proportional-sized circles in the map, Fig.~\ref{fig:maptrix-no-crossing}. Choropleth maps were also investigated, but as the scale of the total and single flows could be very different multiple colour schemes would be needed. In addition to the proportional circles, bar charts were added to help the reader to follow the line (e.g. from large circle to large bar) between map and matrix and to emphasise total in and out flows.

The design is also well suited to showing flow between different countries, as shown in Fig.~\ref{fig:maptrix-no-crossing}(b). However, for showing flow within a single country the asymmetrical ordering of rows and columns in the OD matrix can be confusing. A consistent ordering for the rows and columns is critical for revealing patterns \cite{Guo:2007gi}.

\begin{figure}
\setlength{\abovecaptionskip}{.1cm}
\setlength{\belowcaptionskip}{-0.4cm}
\centering
  \includegraphics[width=0.95\columnwidth]{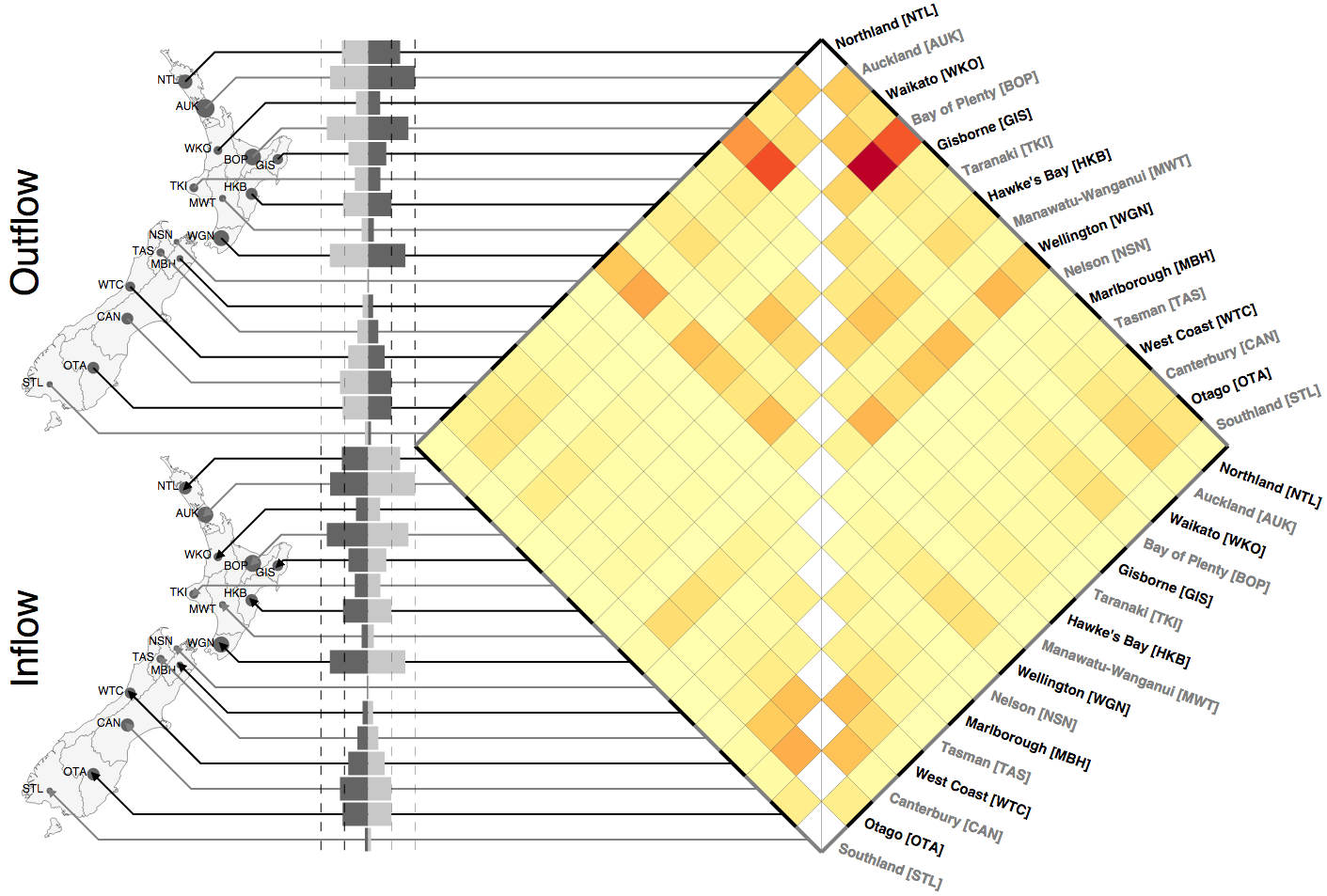}
  \caption{Final design for crossing-free leader lines connecting maps and OD matrix with identical column and row ordering.}\label{fig:maptrix-final}
\end{figure}

Our final design---shown in Fig.~\ref{fig:maptrix-final}---permits the same ordering to be used for rows and columns by rotating the OD matrix. The destination map is placed under the origin map and the OD matrix is rotated to allow both symmetric ordering and crossing-free leader lines to the maps. An additional advantage of the rotated matrix is that the labels are easier to read. To utilise the additional space and aid leader line connection, the bar charts showing total in/out flow are centred on the leader-lines. We also add total inflow and outflow to both bar charts, differentiated by colour, to allow net flow to be easily determined. Instead of the large arrow, we use the darker bar charts to indicate direction of flow. 

\subsection{Algorithm for Leader Line Placement}
\label{sub:maptrix-algorithm}

When connecting sites in the maps with rows and columns in the OD matrix we  would like: 
(1) connection lines to be crossing-free; 
(2) adjacent connection lines to be clearly separated; and
(3) clear separation between lines and map locations (\emph{sites}) to avoid ambiguity.

\vspace{1pt}
\noindent
\begin{minipage}{.76\columnwidth}
\quad The starting point for our algorithm is the one-sided boundary labelling method of Bekos \emph{et al.}~\cite{Bekos:2009id} which orders and spaces labels evenly at one side of the figure. 
The model by Bekos \emph{et al.}\ produces crossing-free, minimal length leaders, each with a diagonal segment of uniform gradient. 
However, while Bekos \emph{et al.}\ can ensure no crossings, their method cannot ensure adequate separation
 between leaders and connection sites or other leaders which may lead to serious ambiguity (see right).

\quad Fortunately, with the MapTrix visualisation we are showing flows between areal regions within which there is typically some freedom to move the connection site of the leader. This means that in a second stage of the layout we can fine-tune the connection site placement so as to increase the separation between leader lines. We use a quadratic program to do this.
\end{minipage}
\begin{minipage}{1.8cm}

\vspace{0mm}
\small
\fontfamily{phv}\selectfont
\centerline{Line through site}
\vspace{1mm}
\includegraphics[height=1.5cm]{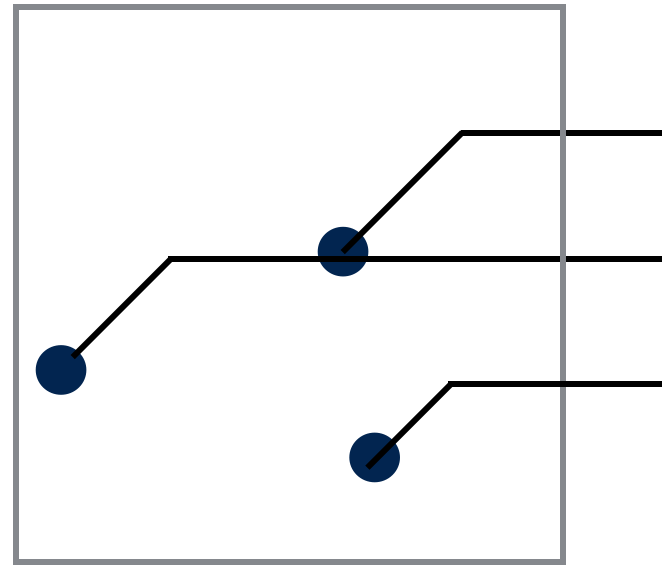}
\centerline{Lines too close}
\vspace{1mm}
\includegraphics[height=1.5cm]{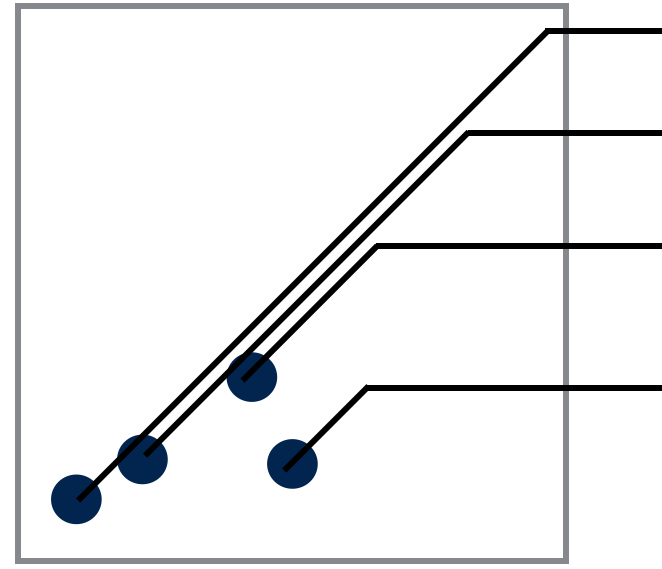}
\centerline{Line overlap}
\vspace{1mm}
\includegraphics[height=1.5cm]{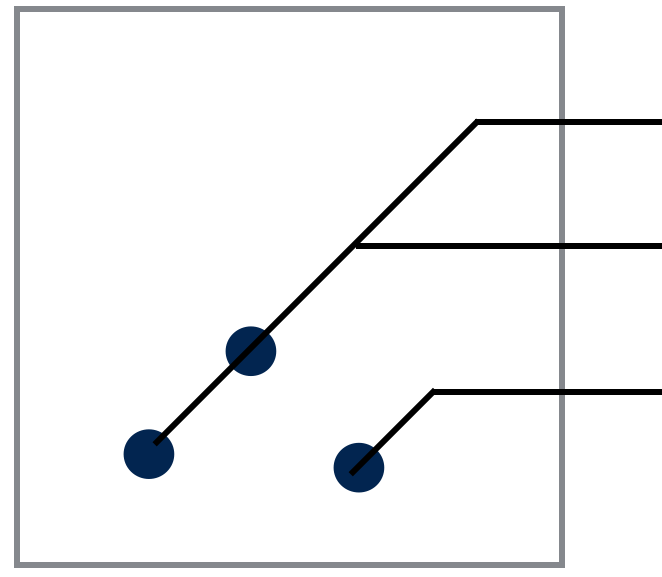}
\vspace{0mm}
\end{minipage}
We associate penalties with close leader-line segments and displacement of connection points from their initial position. We define hard linear constraints to preserve the ordering of leaders and keep the connection points inside their state boundaries.

Input to the Bekos \emph{et al.}\ one-sided boundary labelling is a connection site for each map location, typically at the centre of a region.  The output is a label ordering permitting crossing-free connection to map locations. That is, for $n$ sites $c_{xi}, c_{yi}, 1\le i \le n$ we have an ordering such that the leaders for sites $i$ and $i+1$ are adjacent and crossing free.  There are two types of leaders: those with diagonals pointing upward from the sites and those with downward diagonals. 

The quadratic program to reposition connection sites to achieve good leader separation is as follows.
Let $l_{xi}, l_{yi}$ be variables for leader connection coordinates.
The first set of goal terms penalise displacement of connection sites from their initial position:
$$
  \mathit{PCentre} = \sum_{i=1}^n (l_{xi} - c_{xi})^{2} + (l_{yi} - c_{yi})^{2}
$$

Inside each state boundary we find a rectangle in which the connection site can be safely positioned. Ideally, in order to maximise freedom in placing the connection site, this should be a rectangle with maximal width and height centered around the initial site position. We use a simple heuristic to find such a rectangle. We start from the initial

\noindent
\begin{minipage}{.52\columnwidth}
position of each location and grow a rectangle within the state boundary using binary search, alternating between growing the width and height.  This gives us, for each region $i$ the rectangle with upper-left corner $bu_xi, bu_yi$ and bottom-right corner $bb_xi, bb_yi$. However, this rectangle may permit the connection
\end{minipage}
\begin{minipage}{4.1cm}
\centering
\vspace{0mm}
\small
\fontfamily{phv}\selectfont
Bounding Box Constraint\\
\vspace{1mm}
\includegraphics[width=4cm]{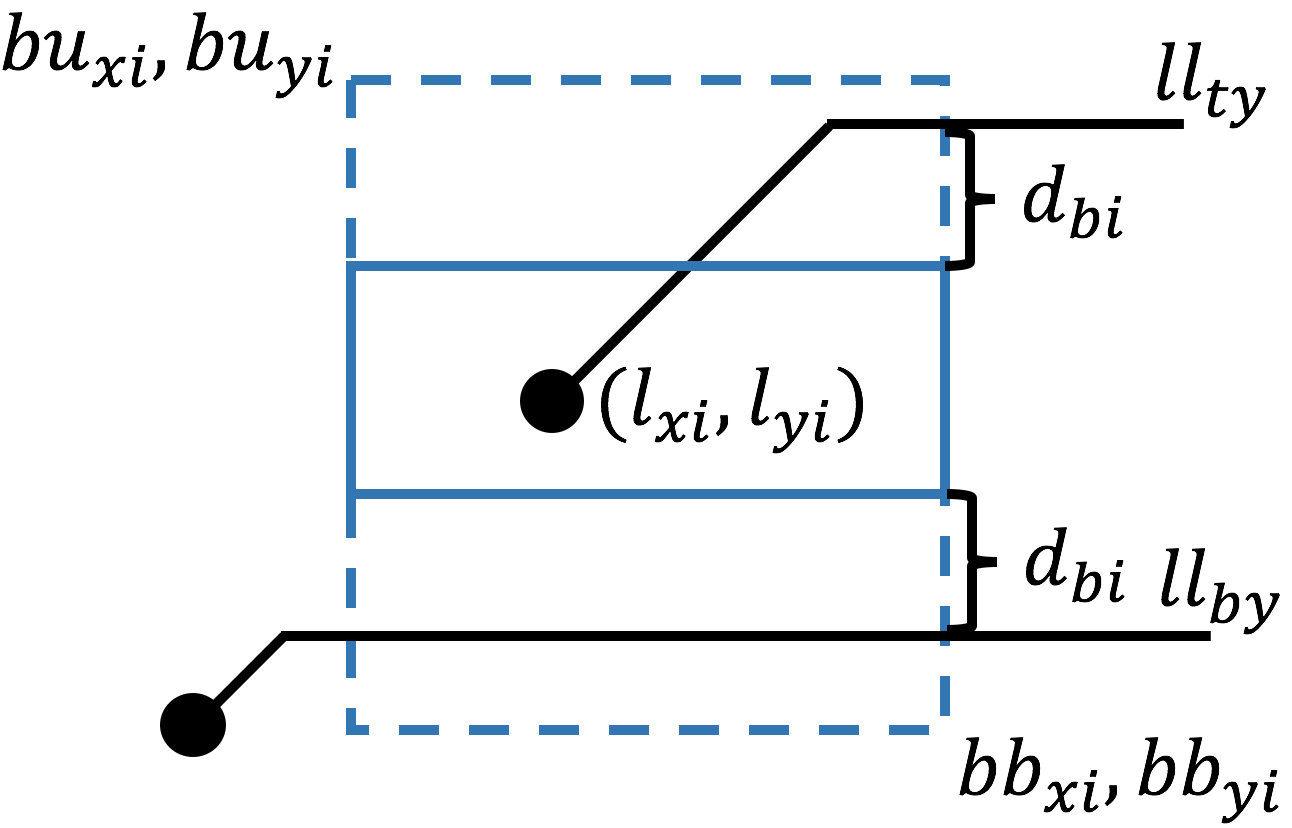}
\vspace{0mm}
\end{minipage}
site to cross another leader line and introduce a crossing as shown in \emph{Line through site}.
Thus, we prune the rectangle to ensure the site remains a minimum distance from all other leader lines ($d_{bi}$).

Constraints keep the leader connections inside the (pruned) rectangle boundaries:
  \begin{equation}
    \label{eq:maptrix-algorithm-within}
    \begin{array}{rcl}
      l_{xi} > bu_{xi}, l_{xi} < bb_{xi}\quad \wedge \quad  l_{yi} > ll_{by} + d_{bi}, l_{yi} < ll_{ty} - d_{bi}
    \end{array}
  \end{equation}

\noindent
\begin{minipage}{.67\columnwidth}
\quad 
The second part of the quadratic model aims to increase the separation between leader lines without introducing crossings. The Bekos \emph{et al.} Algorithm produces alternating bands of leader lines with upward or downward bends. Furthermore, there is horizontal separation between each pair of bands. To ensure that we do not introduce overlap between leaders in two adjacent bands we simply add a separating line  between the adjacent bands (red dashed line $l_c$ in \emph{Center Distance Constraint}) and add a constraint to ensure sites in each band maintain at least a certain distance above or below this boundary line ($d_{lc}$). 

\quad We now consider the case of adjacent leaders in the same band. The distance between adjacent, similarly oriented leader line diagonals (in \emph{Line Distance Constraint}) is given by
\end{minipage}
\begin{minipage}{2.8cm}
\centering
\vspace{0mm}
\small
\fontfamily{phv}\selectfont
Center Distance Constraint\\
\vspace{1mm}
\includegraphics[width=2.7cm]{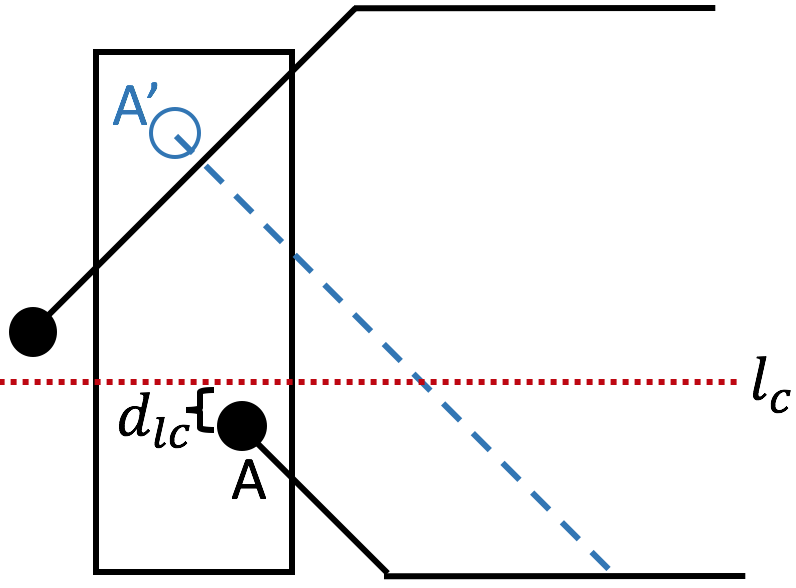}
Line Distance Constraint\\
\vspace{1mm}
\includegraphics[width=2.7cm]{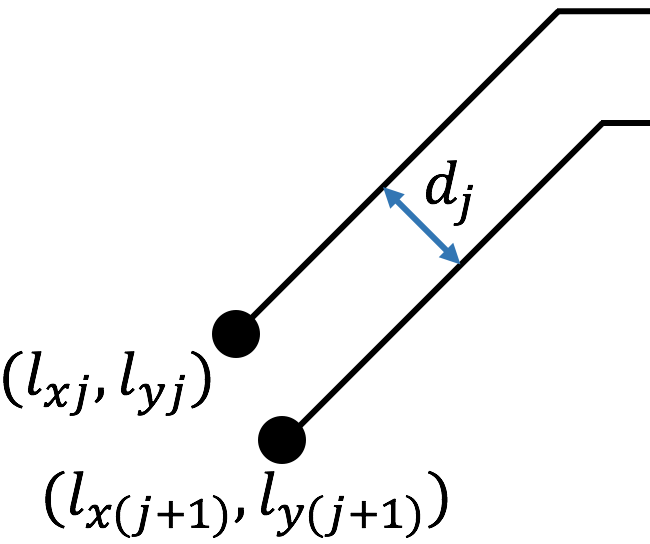}
\end{minipage}
\vspace{1pt}

\begin{equation}
  \label{eq:maptrix-algorithm-distance}
  d_j = \frac{(k l_{x_{(j+1)}} - l_{y_{(j+1)}} - k l_{xj} + l_{yj})} {\sqrt{k^2+1}}
\end{equation}
where $1 \le j < n$ and $k$ is the gradient of the leader diagonals.  Since $k$ is constant the relationship is linear.  We introduce another variable to our quadratic program for each $d_j$ and the above relation between $d_j$ is added as a hard constraint.
The constraint:
\begin{equation}
\label{eq:maptrix-algorithm-ordering}
d_j > 0
\end{equation}
preserves the ordering of parallel leader lines ensuring they remain crossing free.  
A final set of penalty terms encourages equal separation between  adjacent leader lines:
$$
  \mathit{PSep} = \sum_{j=1}^{n-1} (d_j - D)^2
$$
\noindent where $D$ is the maximum initial separation between adjacent leader diagonals output by the Bekos \emph{et al.}\ algorithm. 

The full quadratic goal is $\mathit{PCentre} + w(\mathit{PSep})$ where the weight $w \ge 0$ can be varied to trade-off displacement of connection sites and equal separation of leader diagonals.   To obtain connection sites with good separation we minimise this goal subject to the linear constraints of Equ. \ref{eq:maptrix-algorithm-within}, \ref{eq:maptrix-algorithm-distance} and \ref{eq:maptrix-algorithm-ordering}.  Since the number of variables and constraints is linear in the number of input regions, solving this quadratic program with a standard solver is very fast.  Placement of hundreds of connection sites takes a fraction of a second on a standard computer.

\section{Study 1}
We conducted an on-line user study to evaluate MapTrix  and to compare it  with two alternative visualisation methods; a flow map using bundling and the OD map by Wood et al.~\cite{Wood:2010be}.  We chose these methods because flow maps are the most common visualisation for showing flow while OD maps are an alternative approach to enhance the OD matrix with a geographic embedding. 

We aimed to test the usability of the three methods with respect to various tasks as described in Sec.~\ref{sec:study1apparatus}. We consider user preferences as well as task performance in terms of response time and accuracy. 
This first study considers only static representations. We begin to consider basic interactions in Study 2, Sec.~\ref{sec:redesign}.

\subsection{Bundled Flow Map Design \& Implementation}
\label{sec-bfmd}
\begin{figure}
\includegraphics[width=0.26\columnwidth]{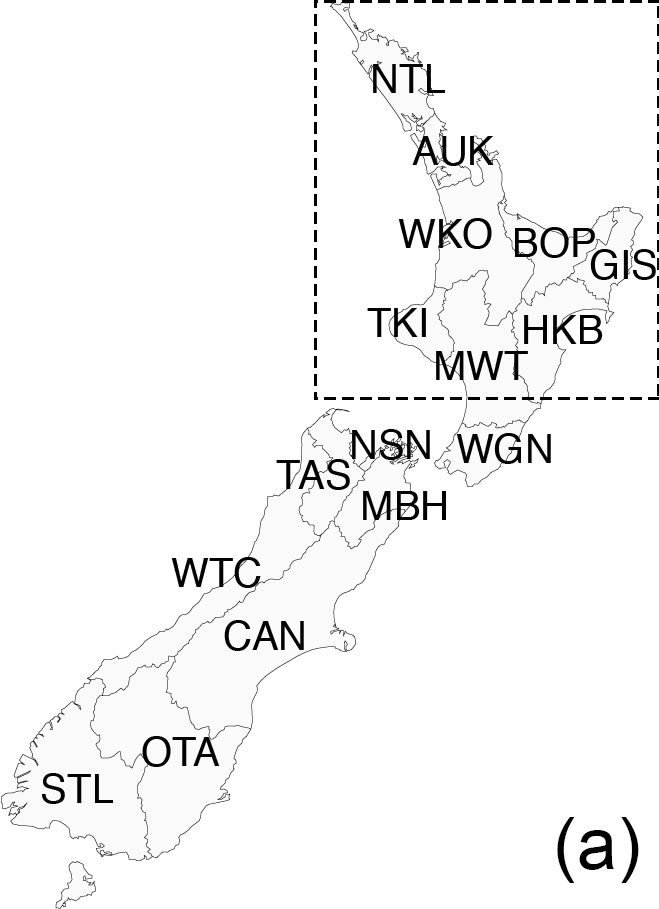}
\includegraphics[width=0.36\columnwidth]{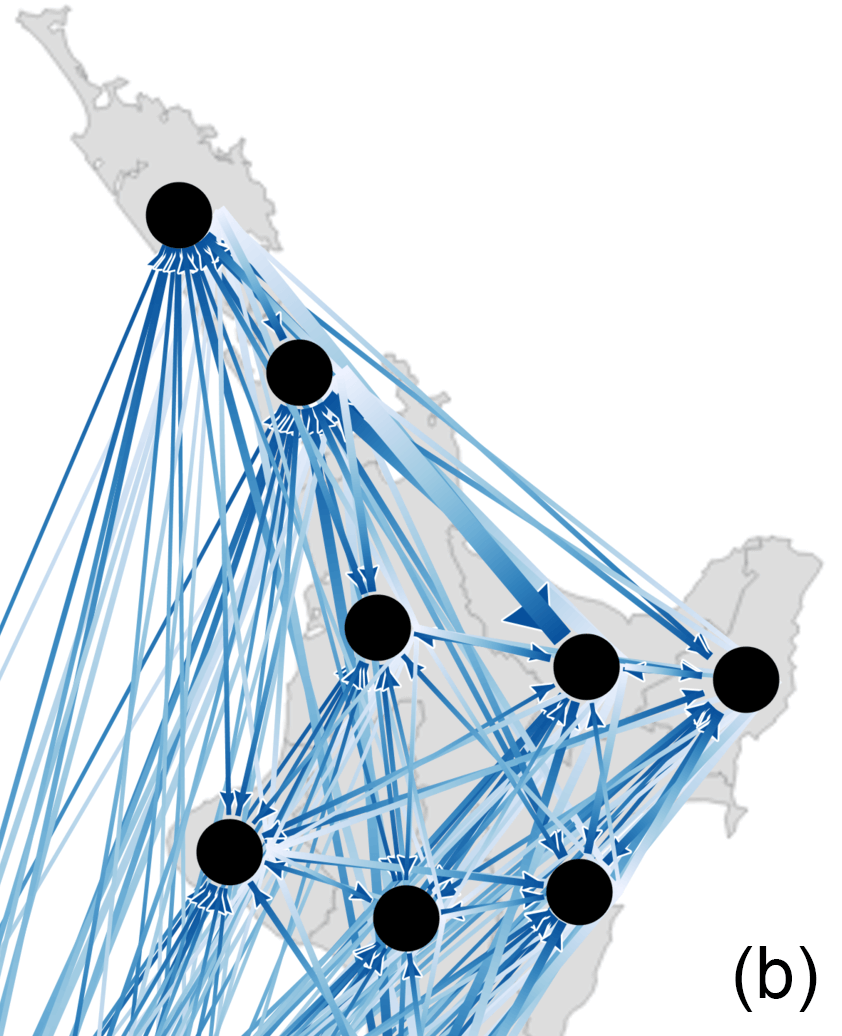}
\includegraphics[width=0.36\columnwidth]{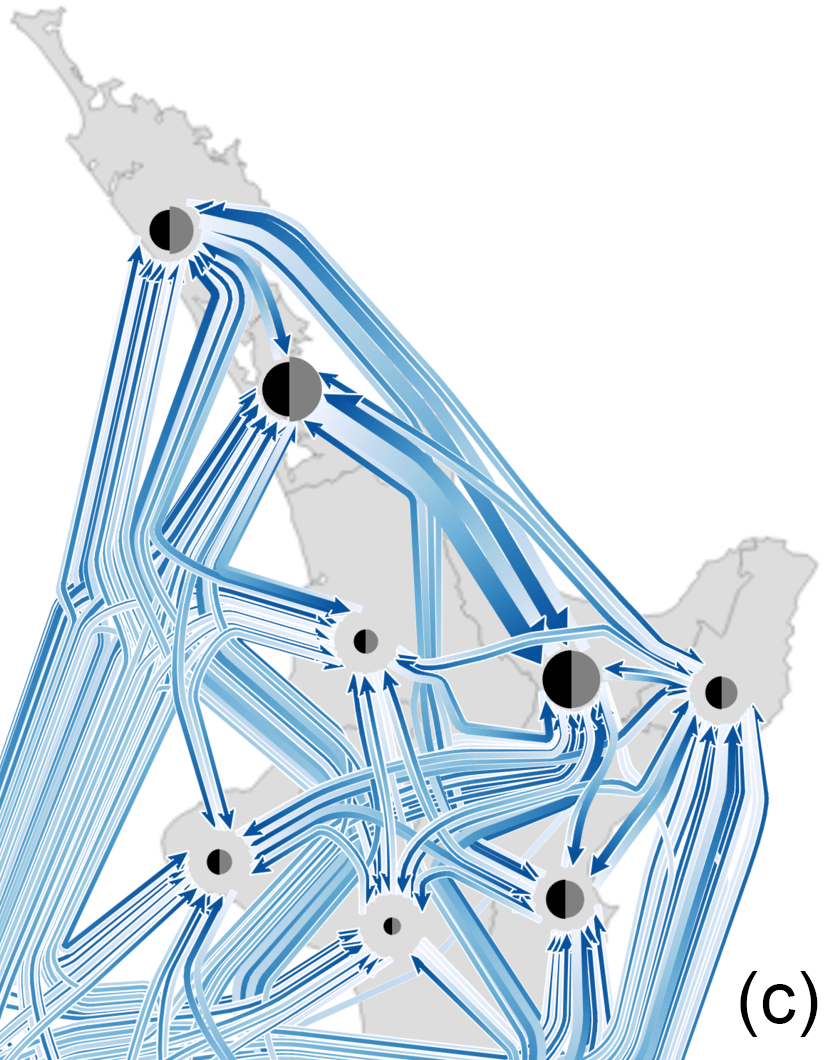}
\caption{Flow map design decisions. Arrow thickness indicates magnitude of flow. Arrow head and colour gradient shows direction. (a) Zoom into a sub region of NZ to have a detailed view; (b) Straight lines with full black circles for locations; and (c) Bundled lines with half circles for locations and magnitude of total in (black) and out (grey) flow.} 
 	\label{fig:bundling-final} 
\end{figure}

We searched for a flow map design solution which could minimise data occlusion by reducing overlap without removing individual flows such as through flow aggregation.  There are a couple of recent edge bundling methods that are able to neatly offset individual edges within bundles \cite{Bouts:2015ip,pupyrev2012edge}.
We adopt the method by Pupyrev \emph{et al.}~\cite{pupyrev2012edge} which groups edges on shared paths that are centred between obstacles.  It then neatly offsets the curves so that all are visible and uses a heuristic to minimise crossings as lines join and leave the bundles. To demonstrate the reduction of line overlap Fig.~\ref{fig:bundling-final} shows straight and bundled arrows.

\begin{table*}
\scriptsize
\setlength{\abovecaptionskip}{.1cm}
\setlength{\belowcaptionskip}{-0.8cm}
  \centering
\begin{tabular}{ | C{0.075\textwidth} | L{0.05\textwidth} | L{0.44\textwidth} | L{0.34\textwidth} | }
    \hline
    Group & Abbr & Description & Example \\
    \hline
    \multirow{2}{*}{\tabincell{c}{Total\\Flow}} 
    	& TFI & \textit{\underline{\textbf{I}}dentify} two total in/out flows for two named locations and compare their magnitude. & Comparing the two locations QLD and TAS, which has the greater total inflow? \\ \cline{2-4}
 		& TFS & \textit{\underline{\textbf{S}}earch} for the largest/smallest total in/out flow. & Which state has the largest total outflow? \\ 
 		\hline
 	\multirow{3}{*}[-5pt]{\tabincell{c}{Single\\Flow}} 
 		& SFI & \textit{\underline{\textbf{I}}dentify} two single flows between named locations and compare their magnitude. & For the two flows from WA to ACT and TAS to SA, which is greater? \\ \cline{2-4}
 		& SFSo & \textit{\underline{\textbf{S}}earch} for the greatest single in/out flow for \textit{\underline{\textbf{o}}ne} named location. & ACT receives the largest single flow from which state? \\ \cline{2-4}
 		& SFSm & \textit{\underline{\textbf{S}}earch} for the largest single flow across all (\textit{\underline{\textbf{m}}any}) locations. & Which is the largest single flow? \\ 
 		\hline
 	\tabincell{c}{Regional\\Flow} & RF & if the flow is predominantly within the \textit{\underline{\textbf{r}}egions} or among the \textit{\underline{\textbf{r}}egions}. & Using the regions A and B defined in the above map, is the flow predominantly within A or B? \\
 	\hline
  \end{tabular}
  \caption{Task description, abbreviations and examples questions from AU}~\label{tab:tasks}
\end{table*}

We investigated the use of colour together with arrow size to indicate magnitude of flow. We found that due to line occlusion around arrowheads the flow direction was often difficult to determine. Following Holten \emph{et al.}\cite{holten2011extended}, we therefore decided to encode line direction using colour gradient. The darker section of the line shows inflow direction, while the lighter section depicts outflow as shown in Fig.~\ref{fig:bundling-final}(c). The continuous blues colour scheme from colorbrewer is used \cite{Harrower:2003jm}. We used a different colour scheme to the MapTrix flow data to limit confusion.  The key aspect of using a continuous gradient from source to target is that the directionality of the line can be understood at any part of the line, so the reader does not need to follow the line to find an arrow head. 
 Note that, since we use line width to encode flow magnitude, the tapered line representation advanced by Holten \emph{et al.} would not work in this situation. 

To embed the information of total in/out flows we use proportional circle sizes. Unlike MapTrix where two maps are available, the bundled map has only one location point. We therefore replaced the solid black circle for each location as shown in Fig.~\ref{fig:bundling-final}(b) with two half circles as shown in Fig.~\ref{fig:bundling-final}(c). The left half circle in black indicates total inflow, while the right half circle in grey shows total outflow. 

\subsection{OD Map Design \& Implementation}
OD Maps preserve the geographical aspects of OD matrices without including lines or arrows and introducing occlusion. 
Having discussed OD Map implementation with the authors~\cite{kelly:2013,Wood:2008,Wood:2010be} we manually created grid layouts for the necessary countries to ensure the grid structure was as intuitive and as similar to the country shape as possible, as shown for Germany in Fig.~\ref{fig:ODmapLayout}. We used the same colour scheme as shown in the MapTrix matrix for the flow data and slightly modified the Wood \emph{et al.}\ OD map design~\cite{Wood:2010be} to include a proportional circle at the associated origin or destination cell of the small multiple to show the total in/out flow for each location. We also show both the OD map for outflows and the reverse `DO' map for inflows, to allow for two way comparison. Fig.~\ref{subfig:grid-au} shows the dual OD/DO Map visualisation shown in the study for Australia.

\subsection{Apparatus \& Materials}
\label{sec:study1apparatus}
\yy{
	In \textbf{summary review}; R3;	describe more precisely the scope actually covered by the studies.
}
\rev{
	R3: Could you state what are the specific goals and the hypothesis of the study?
}
\rev{
	R3: You have asked participants about their background knowledge. Did you try to analyze the data? Are map experts more successful at performing your tasks? Do your conclusions hold across all the groups?
}

In order to test how the visualisations perform for different numbers of locations we investigated their use for different countries. We decided to use real rather than fictional countries and locations  to implicitly emphasize the use of such visualisations for common commodity flows such as population migration. This also allowed us to explore the possible impact of prior knowledge of geography on performance.

\rev{
	R1: it would have been useful to explain how the authors identified the experiment tasks that were used.
}
\rev{
	R3: a little concerned about the chosen tasks.
}

\noindent\textbf{Tasks}
We identified a variety of tasks that commodity flow visualisations should support by reviewing the geographical visualisation literature~\cite{Andrienko:2006, Ali:2013tg, Tobon:2005}. For single and total flows we are mainly interested in flows from a given target location(s), or identifying which location(s) corresponds to a given characteristic. These tend to be lookup or comparison tasks which may refer to identifying and comparing total flow (TF) values of 1, 2 or many locations, or single flows (SF) between 2 or many locations. A further important task involves determine the geographical or regional distribution of the flow (RF). This involves identifying if flow is predominantly within a certain area on the map or between two different areas. We designed our questions of the study into the following six task categories: \textit{TFI, TFS, SFI, SFSo, SFSm and RF}. These are defined along with examples of exact questions in Table~\ref{tab:tasks}. 

\noindent\textbf{Countries and Datasets}
To represent actual commodity flow data, we created synthetic datasets based on real internal population migration data. The first country we chose was Australia (AU) as it has a large spacious country shape with relatively few federal states (and territories). With 8 states there are only $8 \times 8$ between state flows to present. The original dataset for AU is based on 2013-14 internal migration for AU\footnote{http://stat.abs.gov.au//Index.aspx?QueryId=1233}. 

To investigate larger number of flows, Germany (DE) was chosen as a comparison as it again has a large and spacious shape but double the number of federal states and therefore $16 \times 16$ individual flows. For DE between state migration was not openly available so we allocated data from USA internal migration data from 2009-10\footnote{https://www.census.gov/hhes/migration/data/acs/state-to-state.html}. 

A third country, New Zealand (NZ), was chosen to allow us to investigate the effect of country shape. It has the same number of national states as DE but is more elongated. The original NZ dataset is based on regional migration from 2001--06\footnote{http://www.stats.govt.nz/browse\_for\_stats/population/Migration}. 

During our pilot sessions we also investigated countries with larger number of locations, including the United States of America (US) with $51 \times 51$ flows. This number of flows was found to be too confusing and difficult for users, particularly for the bundled flow map design. We therefore removed US from the first study (but used it in the second study Sec.~\ref{sec:redesign}). 

In order to train the participants we introduce the problem and explain the visualisations using the United Kingdom. It has a distinguishable country shape and only four national states (in this case countries) so therefore only $4 \times 4$ flows. The training is provided as supplementary material.

\rev{
	R1: The rationale for real geography is not strong, particularly from an experiment design perspective. The rationale for synthetic data is logical, since it enables them to have different data for every participant and the participants could not rely upon memory to answer. A bit more complete explanation of this would help.
}
For each case we minimise the effect of the data on the study results by randomising the source and destination of the original dataset in order to ensure each question has different data and participants must read the data every task. For Task \emph{RF} (see Table~\ref{tab:tasks}) we ensured that data was different but that the spatial pattern remained. We therefore ensured that the data for analysing flow between/within regions had a definite answer. Sometimes there was a near second best answer.  We experimented with treating these \emph{Almost Correct} responses as correct and as incorrect: this had virtually no impact on the analysis results. In the analysis presented here we give them half points.

\rev{
	R1: The “almost correct” category is odd – I do not believe I have seen any prior study have such a category. Thus, while I am not suggesting that it must be dropped, I do think it requires somewhat more explanation to document why it is included and what the implications of including (or excluding) it are.
}

\noindent\textbf{Procedure}
\label{sub:surveyprocedure}
The structure of the study was slightly amended following the pilot study as the study took too long with all three countries. As all tasks were shown to be important to the analysis we chose to split the countries so each participant was asked questions about only one pair of countries (AU-NZ; AU-DE; NZ-DE). The choice of country pair was counterbalanced. After receiving information about the study through the explanatory statement and agreeing to the consent form the study took the following structure: 
\begin{enumerate}[leftmargin=4mm, labelsep=1mm]

\item Background knowledge: participants were asked about their prior experience using maps -- rarely use, navigation only or often use maps to read statistical information -- and their knowledge of the administrative structure of their pair of countries; 

\item Training: participants were given an overview of the problem and explanation of each of the visualisation methods. Upon finishing the training for each method the participant was showed two sample questions with the answers and explanation. They were then asked to answer another two questions to verify that they understood the method. The training order was counterbalanced.

\item 
Tasks: participants were asked to answer $36 = 6 \times 3 \times 2$ questions: one for each kind of task for each of the three visualisation methods and each of the two countries.  Question order was randomised.

\item
Ranking and Feedback: participants were asked to rank the three visualisation methods in terms of visual design and in terms of effectiveness of reading information for each of the two countries that had been shown. They also had the opportunity to comment on the strengths and weaknesses of each visualisation method.
\end{enumerate}

\noindent\textbf{Participants}
\label{sub:surveyparticipants}
To attract a range of skill-levels amongst participants the study was advertised at Monash University (Australia) using a university-wide bulletin and through email lists at Microsoft Research (USA), HafenCity University (Germany) and two international map visualisation lists of GeoVis and CogVis. Three \$50 gift cards were offered as an incentive, where participants could optionally provide their contact details and be placed in the prize draw. 

\rev{
	R4: subjects in the user study might not be representative of intended users
}
\rev{
	R1: I remain a bit skeptical about the extent to which these results (with an open call to university people) are both generalizable and applicable to the primary target audience of large data innovators. The participants in the study are probably not representative of the scientists that much of the paper seems to suggest will be primary users.
}

In total we had 62 complete responses, with an equal split of country pairs -- 20 AU-DE, 21 AU-NZ and 21 NZ-DE. Of these 2 participants were excluded from the final analysis due to the exceedingly quick completion time of 5m and an average task time of 8s. Upon analysis we also trimmed 1\% of response times -- those over 300s/5m -- this removed large outliers ranging from 305s to 3352s. On average the 60 participants spent 39s per task and the entire online study took an average of 51m:52s to complete.

\rev{
	R1: The results would be much more convincing if power statistics were supplied to go with the research results; this would enable readers to judge whether “significant” differences are likely to actually be meaningful.
}
\noindent\textbf{Statistical Analysis Methods}
We consider response time and accuracy for each question. We investigate the effect of the three conditions of visualisation (\emph{Vis}) (these are abbreviated to \emph{BD} for Bundled Flow Map, \emph{OD} for OD Map, \emph{MT} for MapTrix in this section), country and task, and to what degree these conditions differ significantly. 

In our analysis we treat all conditions as being independent. Although question order was randomised, we validated task independence by plotting results against question order. No clear pattern was  evident.

To compare error rates between different conditions we use standard non-parametric statistics~\cite{Andy:2012ds}: For multiple (more than 2) conditions, we use Friedman's ANOVA to check for significance and apply Post hoc tests with Bonferroni correction to compare groups while for two conditions, we use the Wilcoxon signed-rank test. Both tests require the same participants in all conditions so when comparing across countries (3 conditions) we could not directly use Friedman's ANOVA as participants only completed the study for two countries.  Instead we split the results into 3 groups, one for  each pair of countries and used a Wilcoxin signed-rank test for each group.

To compare response time we consider only times for correct and almost correct responses. To test for significance we use a multilevel model for analysing mixed design experiments~\cite{Andy:2012ds}.  Here we breakdown the analysis by each condition and their interactions (\emph{Vis/Country}, \emph{Vis/Task}, \emph{Country/Task} and \emph{Vis/Country/Task}).

For the user preference results we again use Friedman's ANOVA and Post hoc tests to test for significance.

\subsection{Results}
\label{sec:results}

\rev{
    R3: The layout of the figure makes me compare accuracy of a specific vis method across various datasets. However, I would like to compare different methods on the same dataset. For example, it is almost impossible to say whether BD outperforms OD for dataset AU and task RF. Can you ``rotate'' the figures?
}

\begin{figure}
\setlength{\abovecaptionskip}{.1cm}
\setlength{\belowcaptionskip}{-0.4cm}
\centering
    \includegraphics[width=\linewidth]{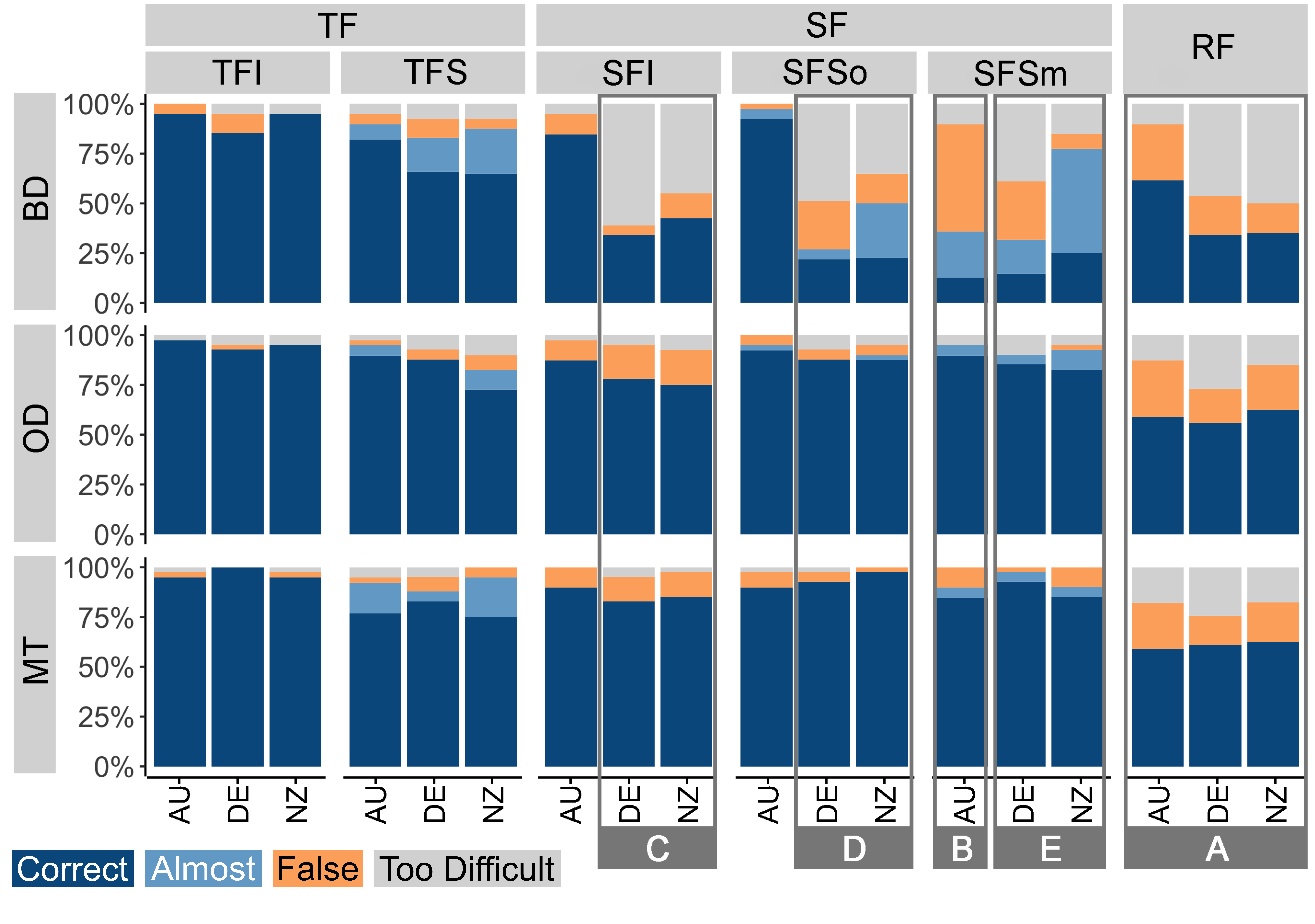}
    \caption{\label{fig:error-1}First study accuracy.  
    Highlights A-E are statistically significant as described in the text.}
\end{figure}

\noindent\textbf{Error Rate}
Responses were in four categories of accuracy: \includegraphics[height=3mm,trim=0mm 2mm 0mm 0mm]{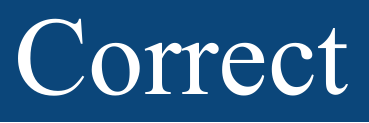}, \includegraphics[height=3mm,trim=0mm 2mm 0mm 0mm]{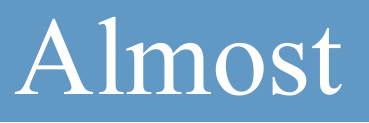}, \includegraphics[height=3mm,trim=0mm 2mm 0mm 0mm]{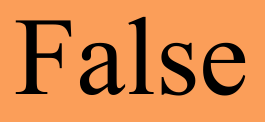} and \includegraphics[height=3mm,trim=0mm 2mm 0mm 0mm]{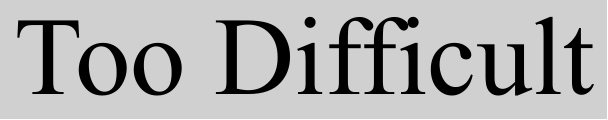},  Fig.~\ref{fig:error-1}. 
We see notable differences in the performance of BD compared to OD and MT, in particular for the SF tasks for the two larger datasets (DE and NZ). All vis methods perform well in the TF tasks, especially TFI. The RF task also shows a far lower accuracy across all vis methods (A in Fig.~\ref{fig:error-1}).

Our smallest dataset (AU) consistently out-performs DE and NZ in almost all tasks for all vis methods.  There is one notable exception (see highlight B: Fig.~\ref{fig:error-1}): 
BD performs far worse for SFSm with 13\% correct + 23\% almost correct, compared to 90\%+5\% for OD and 85\%+5\% for MT. Statistical significance is shown between BD:OD and between BD:MT (both $p< .0001$). No statistical significance is evident between OD:MT. 

The other two countries DE and NZ have the same number of flows. There are some similarities and notable differences when comparing the two sets of results. Most notably, BD is less accurate for SF tasks, see C, D, E in Fig.~\ref{fig:error-1}. 
For all SF tasks using DE and NZ, Wilcoxon signed-rank tests show statistical significance between BD:OD (SFI: $p= .0012$, both SFSo and SFSm $p<.0001$), and between BD:MT ($p$ values: SFI $< .0001$, SFSo $< .0001$ and SFSm $< .0001$).

For SFS(o/m), compared to BD not only does response rate improve using OD and MT (all $p$ for OD:BD \& MT:BD in SFS(o/m) $<.0001$), but the ability to differentiate the dominant answer (i.e. \emph{correct} rather than \emph{almost correct}) is far higher, see D and E in Fig.~\ref{fig:error-1}. 

For TF tasks we observe more similarity between methods. All vis perform well particularly for TFI, with BD performing slightly worse for DE. For TFS we see some differences between vis methods with OD and MT performing better than BD, but no statistical significance is found. 

For RF tasks, not only do we see a difference in performance between all tasks for all vis, but BD performed notably worse than OD and MT (See Fig.~\ref{fig:error-1} A).
Friedman's ANOVA (details Sec.\ \ref{sub:surveyparticipants}) for DE reveals a statistical significance between BD:OD ($p< .0001$) and between BD:MT ($p< .0001$), the same for NZ; between BD:OD ($p=.0108$) and between BD:MT ($p = .0108$). Similar percentages are reported for both DE and NZ, with 34 and 35\% for BD and between 56 and 63 for OD, and 61 and 63\% for MT.  No statistical significance is again found between OD and MT.

\begin{figure}
\setlength{\abovecaptionskip}{.1cm}
\setlength{\belowcaptionskip}{-0.4cm}
\centering
    \includegraphics[width=.5\textwidth]{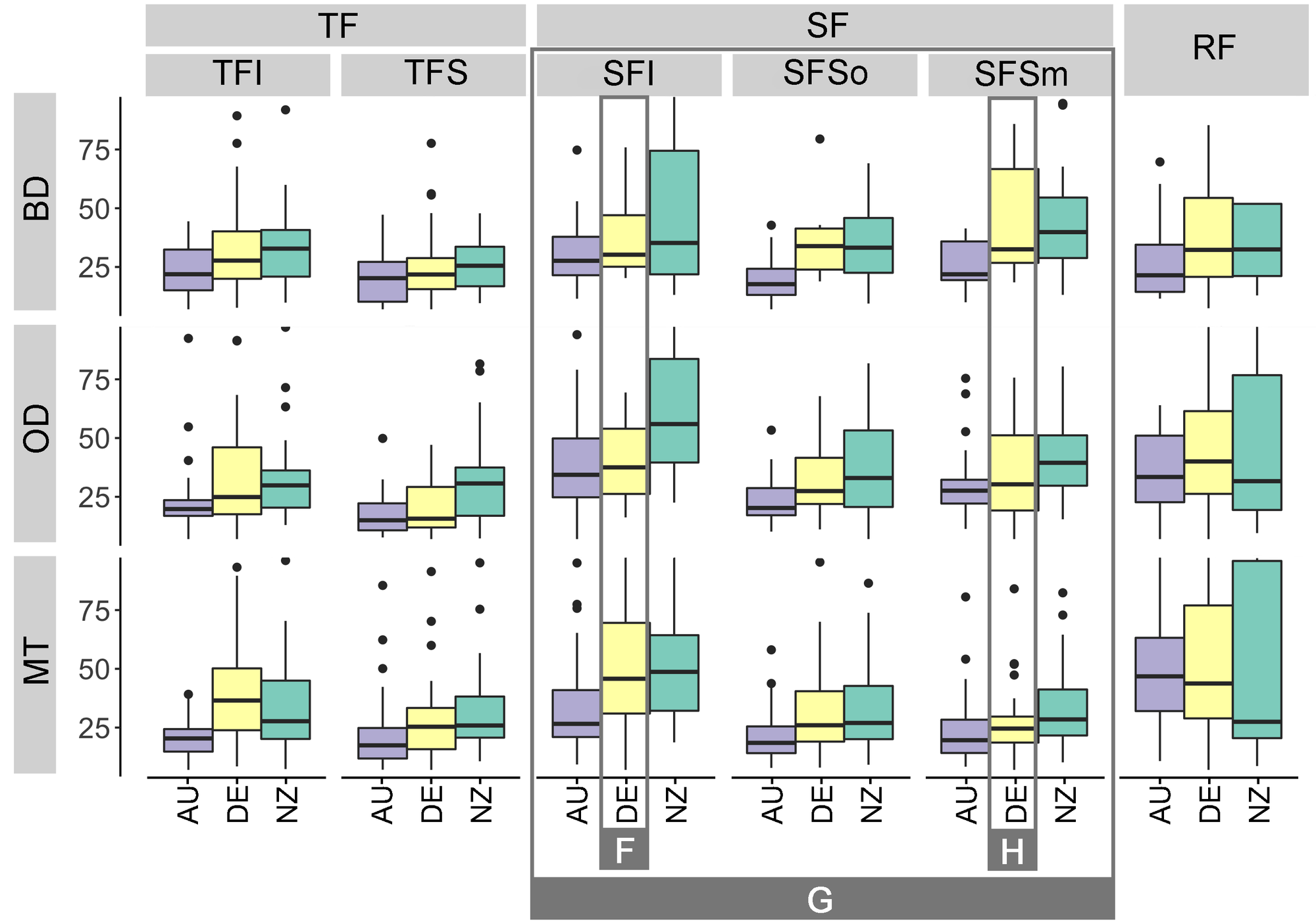}
    \caption{\label{fig:time-1}First study response times in seconds. Highlights F-H are statistically significant.}
\end{figure}

\noindent\textbf{Response Time}
We extract the results of all \emph{Correct} and \emph{Almost Correct} responses (1689 timed responses) from all 2160 responses and plot these for all conditions, as shown in Fig.~\ref{fig:time-1}.  
These box plots, together with multilevel model analysis method, reveal: 

For DE, Task SFI takes increasingly longer from BD, OD and MT (i.e. MT $>$ OD $>$ BD -- see F in Fig.~\ref{fig:time-1}). This is shown to be statistical significant ($p=0.0087$);

For DE, SFSm the trend is the opposite (i.e. MT $<$ OD $<$ BD) -- see G in Fig.~\ref{fig:time-1}. Again, there is a statistical significance ($p=0.0485$);

Although their accuracy is higher, OD and MT took notably more time on RF than BD. MT longer than OD. Correct responses have a wider range for NZ. No statistical significance is found.

Finally, as the size of the dataset increases from AU to DE/NZ we see increasing response time for all tasks, especially for SF tasks (multilevel comparison: DE $>$ AU \& NZ $>$ AU, $p=0.0003$) (see Fig.~\ref{fig:time-1} H). NZ often takes longer than DE.

\noindent
\begin{minipage}{.71\columnwidth} 
\noindent\textbf{User Preference \& Feedback}
Participant ranking of visual design for each of the three methods---by percentage of respondents---is shown here by colour: \includegraphics[height=2.5mm,trim=0mm 3mm 0mm 0mm]{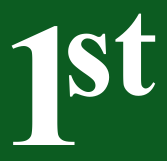}, \includegraphics[height=2.5mm,trim=0mm 3mm 0mm 0mm]{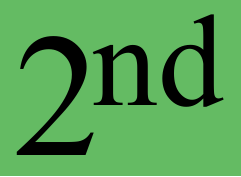} and \includegraphics[height=2.5mm,trim=0mm 3mm 0mm 0mm]{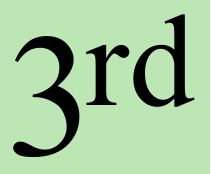}.  The strongest preference is for MT, with almost 50\% of respondents voting MT first place. The other two methods
each received approximately 25\% of votes.  

These differences have statistical significance $\mathrm{OD}>\mathrm{BD}$ with $p=0.0035$, $\mathrm{MT}>\mathrm{OD}$ and $\mathrm{MT}>\mathrm{BD}$ both $p<0.0001$.
Participant ranking of readability is similar and, again, statistically significant with $\mathrm{MT}>\mathrm{OD}>\mathrm{BD}$, all $p<0.0001$.

The final section of the study allowed participants to give feedback on the pros and cons of each design. Qualitative analysis of these comments reveal (overall):
\end{minipage}
\begin{minipage}{2.5cm}
\centering
\vspace{0mm}
\small
\fontfamily{phv}\selectfont
Visual Design Ranking\\
\vspace{1mm}
\includegraphics[width=2.5cm]{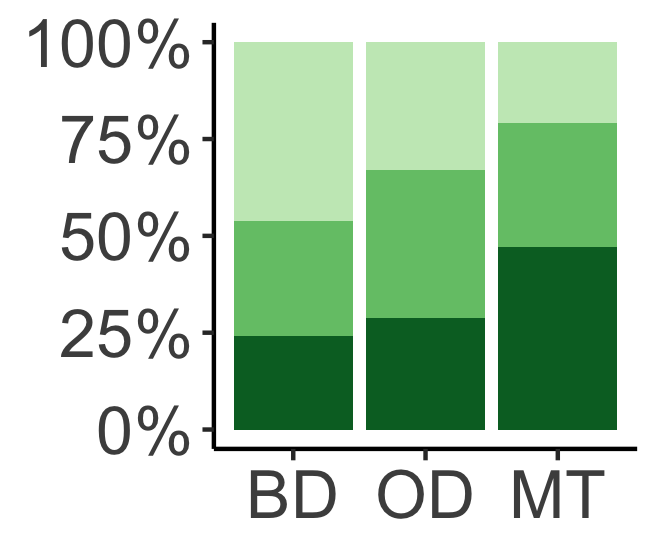}   
Readability Ranking\\
\vspace{1mm}
\includegraphics[width=2.5cm]{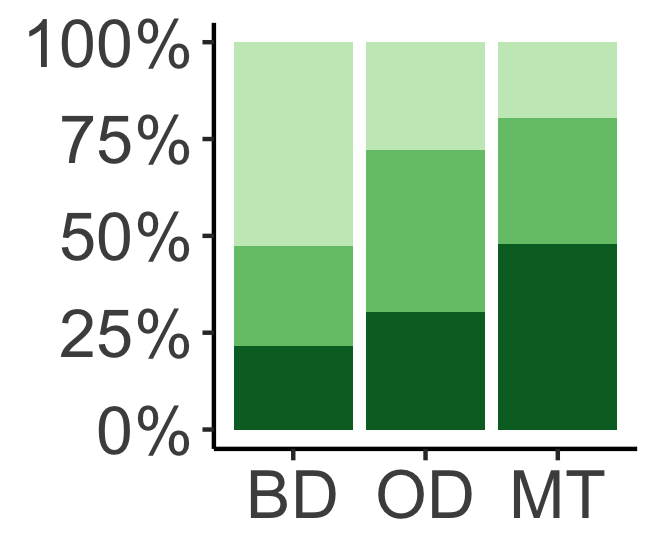}
\vspace{0mm}
\end{minipage}  

\noindent \textbf{BD} was intuitive and familiar: \textit{``it is good for anyone with geographic knowledge and spatial cognition''}. But arrows overlap, arrows are too long, the visualisation does not help when there are many locations and was hard for the RF task: \textit{``Too many locations means many arrows, they occlude, it's hard to see which is which''} and \textit{``Hard to follow arrows over long distances or through intersections... impossible to answer the between or within regions questions''}.

\noindent \textbf{OD} was easy to comprehend and participants often liked the geographical layout. Others found it good for comparison and easy to read the flows. Some also commented on the novelty: \textit{``It is creative and clear''}. Yet, it was also seen as the most unfamiliar and sometimes difficult to comprehend: \textit{``Arrows are missing, I was confused to identify inflow and outflow''}. The small square sizes were also frustrating:  \textit{``the visualisation (grids) can become rather small and more difficult to interpret''}.   

\noindent \textbf{MT} was visually attractive, easy for larger flows and intuitive: \textit{``clear, it is easy to quantify the flows''}. Yet, some found it confusing or unfamiliar.  A few commented that it looked complex: \textit{``It may look complicated but it is the best visualization for information extraction''}. Others found it difficult to follow the lines or read the labels in the matrix, especially with more locations: \textit{``When dataset is large, it becomes difficult to follow the flow''}. Some also commented on there being too much information and there being redundant visual elements (e.g. bars and circles).

\noindent\textbf{Summary}
These results reveal that:
\begin{itemize}
\item AU is the fastest and best performing of all countries. All vis methods are suitable for such small datasets, with the exception of BD for the SFSm task;
\item Error rate worsens with scaling data from AU to NZ/DE, especially for BD for all the SF tasks, where OD and MT out-perform BD with statistical significance; 
\item SFI takes the longest of the SF tasks and on average has the highest error rate;
\item The RF task takes the longest and has the highest error rate compared to all other tasks. All vis methods performed poorly;
\item OD and MT show no significant differences in performance across all conditions;
\item Participants prefer MT for design and readability of information.
\end{itemize}

\noindent
A central design goal of OD and MT is to overcome the problem of occlusion of flows as data increases. 
For the larger datasets (DE and NZ) both OD and MT were significantly better than BD, but there is no significant difference between the two for any condition. User ranking indicates a preference for MT. The fact that BD performs worse is unsurprising given the known problem of overlapping flows; however, the remarkably similar performance of OD and MT is unexpected. We now examine to what extent these methods scale and whether the similarities in task performance continue with increasing scale.
\section{Redesign and Study 2}
\label{sec:redesign}
In this section we concentrate on the scalability of MT and OD. Our second study followed the same structure and participant recruitment method (see Sec.~\ref{sub:surveyprocedure} and ~\ref{sub:surveyparticipants}) as the first, but the countries investigated were amended together with improvements made to the tasks and visual designs.

\noindent\textbf{Data} 
To investigate larger countries than NZ and DE ($16 \times 16$ flows) we wanted to use \emph{The United States of America (US)} with $51 \times 51$ flows as our previous pilot revealed that participants got frustrated with BD for US, but less so with OD or MT. We also chose \emph{China (CN)} with $34 \times 34$ flows as it is almost half way between the two.  For CN the original data set is available for the internal migration from 2005-10~\footnote{http://www.stats.gov.cn/tjsj/pcsj/rkpc/6rp/indexch.htm}. For US we use 2009-10 internal migration data~\footnote{https://www.census.gov/hhes/migration/data/acs/state-to-state.html}. Again we randomised the locations of the data for each question.

\noindent\textbf{Tasks}
In the first study, the RF task was found to be extremely difficult. However, when considering flow in a geographical context it is important to be able to easily compare multiple groups of locations.  In the description of tasks below, we define a \emph{region} as being a collection of locations on the map that are geographically contiguous (adjacent).  
Due to the design choices of both OD and MT the marks corresponding to flows for such regions in the map may not be adjacent in the visualisation. 

For more detailed comparison, we divide the RF task from Study 1 into subtasks related to the adjacency of regions in the visualisation and whether the flow is occurring ``between'' regions or ``within'' a region.  The six subtasks are labelled: RFBb, RFBw, RFBn, RFWb, RFWw, RFWn.  These codes are explained as follows (examples are provided as supplementary material):
\newline
\newline
\noindent Assume two regions A \& B each consisting of multiple contiguous locations.
    Are the flows predominantly

    \textit{\underline{\textbf{B}}etween}: between A to B or B to A?

    \textit{\underline{\textbf{W}}ithin}: within A or B?

\noindent Different adjacency conditions for visuals of regions and locations:

    \textit{\underline{\textbf{b}}etween}: locations within region and regions are adjacent in vis;

    \textit{\underline{\textbf{w}}ithin}: only locations in each region are adjacent in vis;

    \textit{\underline{\textbf{n}}one}: both are not adjacent in vis; \newline

For each question we manually identified appropriate regions for the task for each visualisation method to ensure comparability. These were combined with the same tasks as the first study. In total, participants were asked to answer $44 = 11 (Tasks) \times 2 (Vis) \times 2$ (Country) questions.

\begin{figure}
\setlength{\abovecaptionskip}{.1cm}
\setlength{\belowcaptionskip}{-0.4cm}
\centering
  \includegraphics[width=1.05\columnwidth]{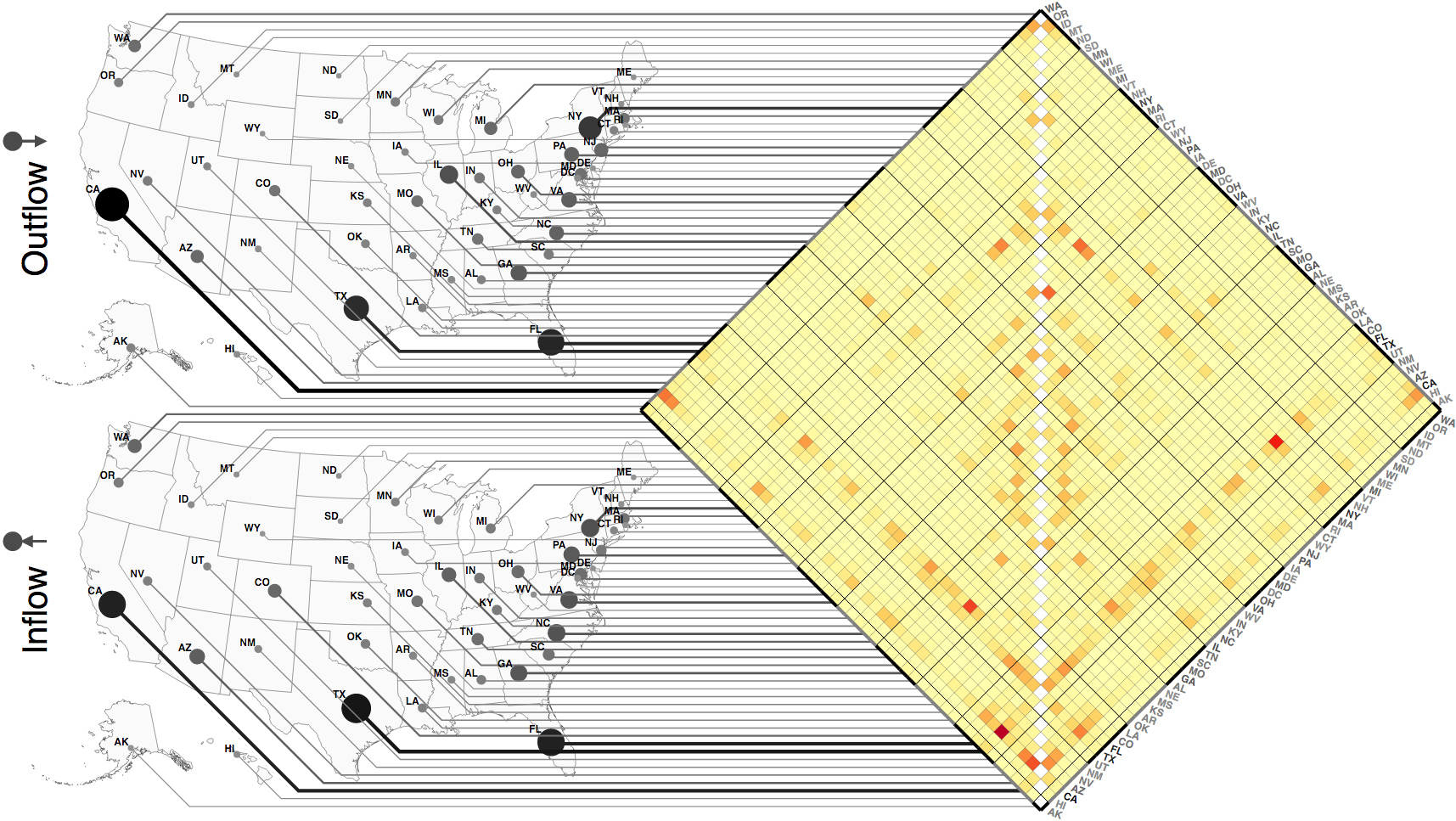}
  \caption{Redesign of MapTrix.}\label{fig:maptrix-redesign}
\end{figure}

\noindent\textbf{Visualisation Redesign}
Feedback and suggestions from the first study and modifications to make it more capable with large dataset, led to some design improvements for both methods explained below.

MT, shown in Fig. \ref{fig:maptrix-redesign}:
\begin{itemize}
    \item Removed TF barcharts to allow more lines, improve tracking and reduce redundancy of information;
    \item Scaled lines thickness and grey shading of line and label proportional to TF circle size and aid line tracking;    
   \item Added separation lines within matrix every 5 rows / columns to aid user tracking;
   \item Minimised overlapping of circles and labels in maps;
  \item Removed full names in matrix. All labels refer to abbreviations;
   \item Removed the arrows in the destination maps to give more space for labels and circles; instead, we  used a destination icon next to the map label. 
\end{itemize}

OD (examples are provided as supplementary material): 
\begin{itemize}
   \item Removed white space to increase grid square size;
   \item Moved and enlarged legend to improve lookups and to allow more space for grid;
   \item Extra care with grid layout to ensure neighbouring regions were adjacent and limited white space -- the downside being that the country is more abstract;%
  \item Increased text size and added label shading to relate to TF and match to the proportional labels in MT;
   \item Added destination icons to indicate direction to match new icons in MT, with multiple arrows in/out compared to only one for MT.
\end{itemize}

\noindent\textbf{Pilot Test and Highlighting}
The first study indicated that the RF task was the most difficult and time consuming across all vis techniques. Our redesign of the RF question to investigate adjacency was intended to investigate this task in more detail; however, pilot testing revealed difficulties.

The RF tasks, although now possible to answer, still took considerable time and were a particular cause of frustration. One participant took over 1h:30m to complete the pilot, with the majority of this time spent manually connecting flows or identifying the squares for the regional tasks. 

To continue to investigate scalability and to allow us to determine whether one visualisation out-performs the other for the aggregation of flows we opted to aid the users in finding the right locations by highlighting them on the OD Map or MapTrix.  Our assumption is that such simple highlighting is easily made available with interaction. We eventually implemented this, see Section \ref{sec-interaction}.
Subsequent pilots revealed much more satisfied users and much faster completion time. 

To encourage participants to think over their answers, instead of showing ``Too difficult'' option straight away we revealed it after 1 minute for every question.

\subsection{Results}
\begin{figure}
\setlength{\abovecaptionskip}{.1cm}
\setlength{\belowcaptionskip}{-0.5cm}
\centering
    \includegraphics[width=\linewidth]{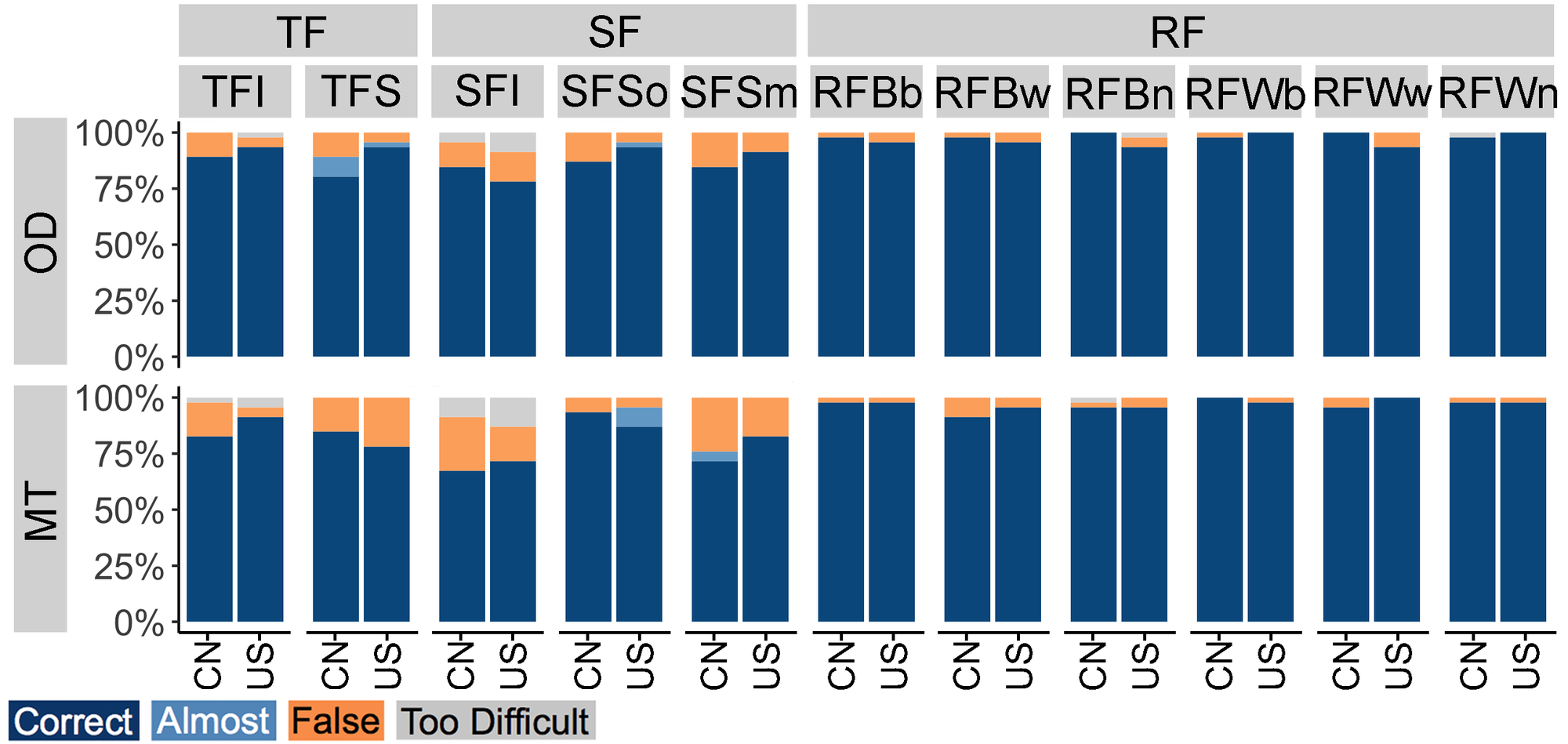}
    \caption{\label{fig:error-2}Second study accuracy.}
\end{figure}

\begin{figure}
\setlength{\abovecaptionskip}{.1cm}
\setlength{\belowcaptionskip}{-0.5cm}
\centering
    \includegraphics[width=.5\textwidth]{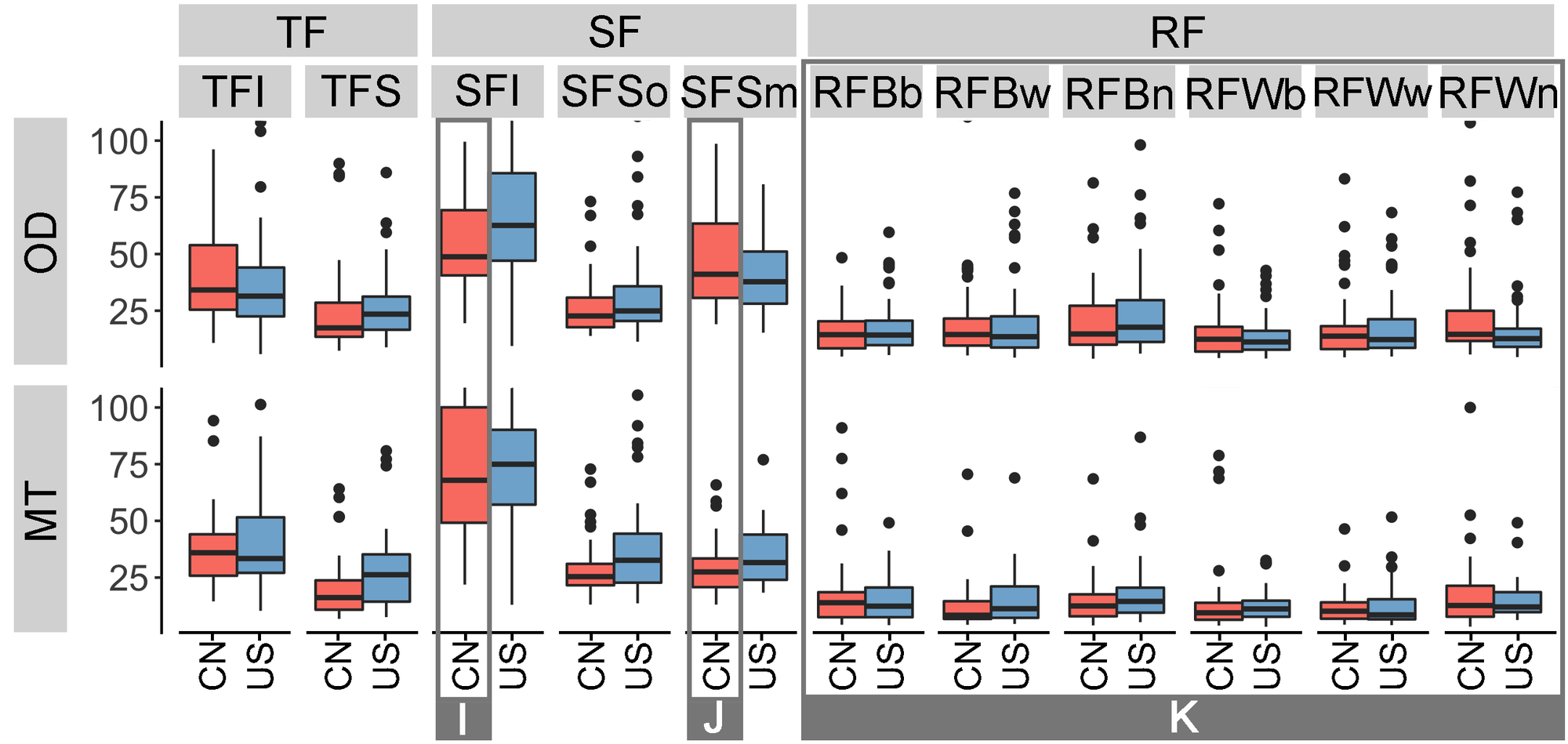}
    \caption{\label{fig:time-2}Second study response times in seconds. Highlights I-K are statistically significant as described in the text.}
\end{figure}

The study had 46 valid responses from an original 49 (3 with impossibly quick responses were excluded). %
On average, individual task completion time was 31.74s and the entire study took 45m:12s. We present the results for error rate and response time in Fig.~\ref{fig:error-2} and Fig.~\ref{fig:time-2}. For the response time analysis, we took the 1861 \emph{correct} and \emph{almost correct} responses from 2024 total responses.

\noindent\textbf{Error Rate}
Fig.~\ref{fig:error-2} shows remarkably similar results across all conditions. No differences are evident in the RF tasks, which all performed very well. Some differences are evident in Fig.~\ref{fig:error-2} between the vis methods for the SF and TFS tasks, but these are not consistent between countries and these differences are not statistically significant. Considering the increase in data flows, it is surprising to see that the results often show an improvement for US over CN. Investigating whether task performance improved with country knowledge, we compare the results for those who claimed good knowledge of the states of US (12 participants) or CN (11) to those who claimed little to no knowledge of the states.  As expected identification tasks (TFI and SFI) increase in speed as well as accuracy for those with knowledge of US, however, only SFI shows an increase for CN. Perhaps the US map is more well known than participants realise, or perhaps it is better suited for these designs.  Feedback from one of the pilot participants suggested that the block shapes of US states helped identification. 

The differences for SF tasks show for CN, OD (82\%) outperformed MT (62\%) in SFI, with slight improvement for both for the US. For CN, OD (82\%) outperformed MT (65\%) in SFSm, with improvements for both for US. This is the only task where one method outperforms the other for both data sets. For CN, MT (91\%) slightly outperformed OD (85\%) in SFSo, but for US the results reduce for MT and increase for OD. 

The final notable difference is for TFS, where for US OD (98\%) outperformed MT (74\%), but for CN results for relatively similar for both vis methods.

\noindent\textbf{Response Time}
In our response time analysis, Fig.~\ref{fig:time-2} shows all conditions. Some notable and statistical significant differences are evident:
\begin{itemize}
  \item In CN for SFI, MT took longer than OD ($p=0.0373$), see I in Fig.~\ref{fig:time-2}, reflecting the increased difficulty indicated through the higher error rate;
  \item In CN for SFSm, OD took longer than MT ($p<0.0001$), see J in Fig.~\ref{fig:time-2}, despite a lower error rate;
   \item With OD, SFSm took longer than SFSo ($p < 0.0001$);
  \item In general, SFI took longer than all other tasks. This is statistically significant for OD in US and MT in both US and CN with ($p=0.0373$).
  \item RF is significantly quicker than the rest ($p=0.0173$). 
 \item RFB[bwn] take significantly longer than RFW[bwn] ($p=0.0404$), see K in Fig.~\ref{fig:time-2}. 
 \item MT is quicker than OD for RF, but not significantly.
\end{itemize}

\noindent\textbf{User Preferences and Feedback}

\noindent
\begin{minipage}{.78\columnwidth}
\noindent
Participant ranking for each of the two methods, by percentage of respondents is shown here by colour: \includegraphics[height=3mm,trim=0mm 3mm 0mm 0mm]{figures/FirstPrefColour}, \includegraphics[height=3mm,trim=0mm 3mm 0mm 0mm]{figures/SecondPrefColour}. The majority of participants ranked MT first for design (63\%).  Compared to the previous study, OD now replaces MT as first for readability (60.9\%). The difference in percentages are marginal and not statistically significant (visual design: $p=0.0641$; readability: $p=0.1228$).  Nevertheless, we investigated whether these rankings relate to participants' knowledge of the country or their map knowledge but found no difference between these groups and the overall ranking. Investigating whether previous experience of these designs affect these rankings, we see only slight differences.
\end{minipage}
\begin{minipage}{1.8cm}
\centering
\vspace{0mm}
\small
\fontfamily{phv}\selectfont
Visual Design Ranking\\
\vspace{1mm}
\includegraphics[width=1.9cm]{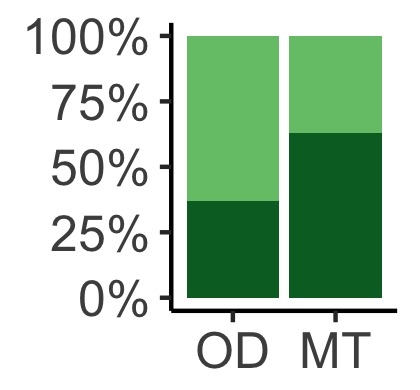}
Readability Ranking\\
\vspace{1mm}
\includegraphics[width=1.9cm]{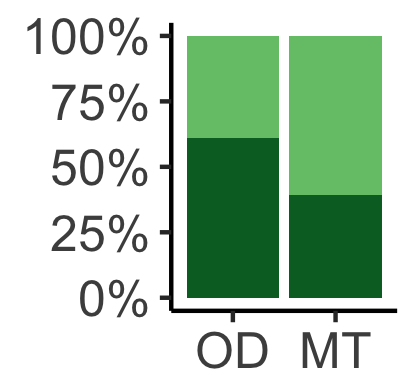}
\vspace{0mm}
\end{minipage}
\noindent
Removing the 6 participants who had participated in the first study from the results we see the rankings for readability of OD increase to 65\%, whilst design remains the same.  

\begin{figure*}
\setlength{\abovecaptionskip}{.1cm}
\setlength{\belowcaptionskip}{-0.5cm}
\centering
\includegraphics[width=0.99\textwidth]{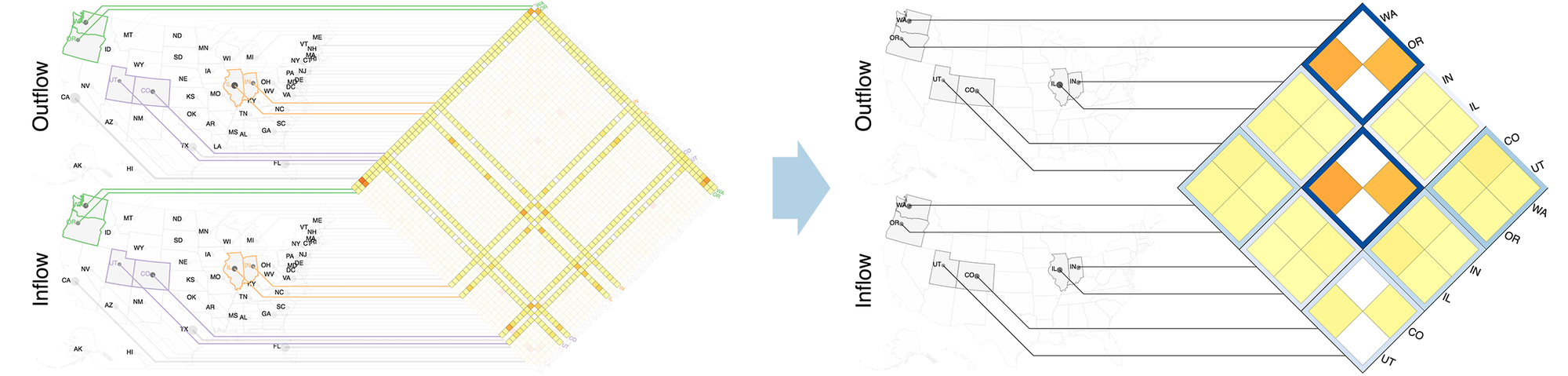} 
\caption{
\label{fig:interaction-region}
A set of contiguous regions can be selected for comparison in a detailed MapTrix view, this also causes a relayout. 
}
\end{figure*}
\begin{figure}
\setlength{\abovecaptionskip}{.1cm}
\setlength{\belowcaptionskip}{-0.4cm}
\centering
  \includegraphics[width=0.96\columnwidth]{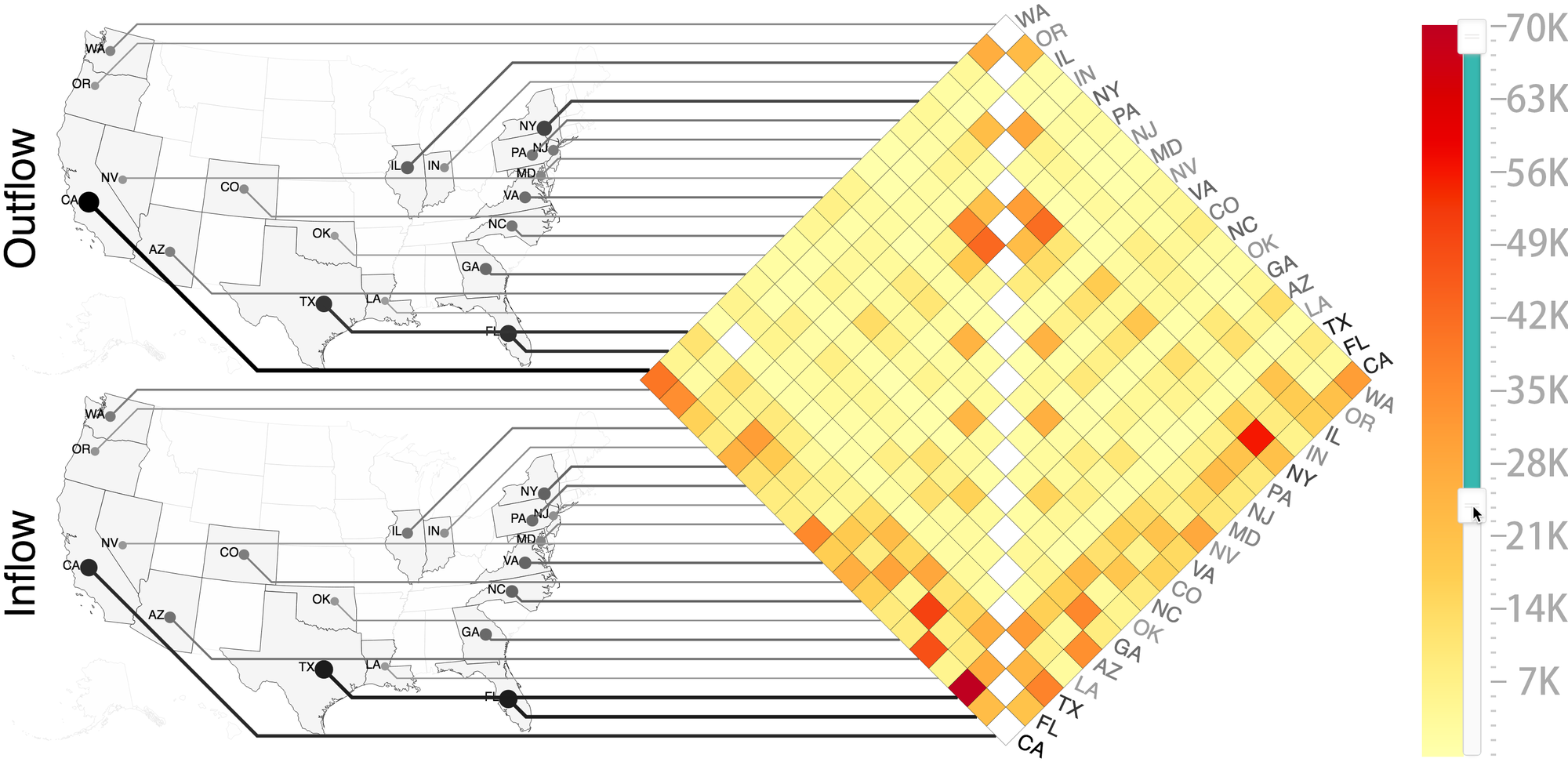}
  \caption{The maptrix display can be limited to show only a certain range of values, this triggers a relayout of the matrix and leaders\label{fig:interaction-label}}
\end{figure}

The qualitative analysis of the feedback quotes again reveals quite conflicting preferences:

\noindent \textbf{OD} was seen as easy to link locations with visualisations. Some participants found it easy to compare single flows: \emph{``OD is easy to find the flow from one location to another without losing your place''} and to find the location names.  Many found the visual elements (grids, cells, circles, labels) far too small. Some disliked the abstract geography: \emph{``losing some geographical reference make locations confusing, given prior map knowledge''}, whilst others recognised that although it can be difficult at first, you can learn the representation: \emph{``you would quickly learn their locations''}.

\noindent \textbf{MT} was found to be familiar because it has real maps and related to the geographical locations. For some the matrix display was also familiar: \emph{``Closer to familiar matrix display. The way of connecting the maps on the left with the rows and columns of the matrix works well.''}. Many participants commented on the difficulty of finding locations, e.g. \emph{``It was too dense with the labels too small to identify the place.''}. Whilst some commented that it was difficult to trace the leader lines and there was a need for a marker: \emph{``Sometimes I had to use a ruler to find the intersection.''}

In general, many participants requested interaction, such as highlighting and selecting. Some noted that locations need to be easier to find in MT, i.e. through reordering the matrix or by allowing text based searching. A few participants also commented that the RF task would be near impossible without the highlighting.

\noindent\textbf{Summary}
The results are consistent with the previous study in that both OD and MT perform similarly. We demonstrate that both methods can scale to data sets containing $51 \times 51$ flows (US). However, the identification of individual and in particular regional flows became much more difficult and time-consuming. SFI takes considerably longer to complete. 

RF took too long, and therefore we aided users with highlighting to simulate possible interaction. Our results using highlighting for all RF tasks are promising for both methods. No clear differences are evident when the adjacency of the regions differ and although the RFB tasks did take longer than the RFW tasks, the accuracy remains very good and response time is relatively low for all RF tasks. 

We do see tasks decreasing in accuracy between the two studies, but in this study, despite both countries being round rather than elongated, we did not find consistency in increased time or error rate with increased numbers of flow -- US outperformed CN for some tasks.

\section{Interaction}
\label{sec-interaction}
\yy{
	In \textbf{summary review}; R1 and R2: discuss interactivity in more detail.
}

We learned from our second study that while MapTrix makes it possible to read a single flow value between a given source and destination in larger datasets it did become more difficult.  For comparing clusters of locations our pilots revealed that highlighting of paths (from origins via leaders and matrix cells to destinations) was essential.  We have constructed a prototype interactive system which allows users to interactively create these highlights through various selection mechanisms \cite{yalong:2016}. 
In particular, the following interactions directly support the indicated tasks from Table \ref{tab:tasks}:
 
 \noindent \textbf{SFSo}--Highlighting of the associated row/column on mouse-hover over a map region, cell, label or leader line.

 \noindent \textbf{TFI, SFI, SFSo}--Mouse-click makes such cell highlighting persist such that multiple flows can be compared simultaneously.

 \noindent \textbf{TFS, SFSm}--The colour key beside the MapTrix is an interactive widget allowing for filtering the MapTrix to a particular range of flow values, Fig.~\ref{fig:interaction-label}.

 \noindent \textbf{RF}--Aggregate selection for Regional Flow comparison tasks, Fig.~\ref{fig:interaction-region}.

The last two interactions both reduce the number of regions shown in the MapTrix and induce a re-layout of the MapTrix and leader lines.  Such re-layout is fast to compute; for the US with 51 locations it is in the order of a few milliseconds.    This dynamic rearrangement, together with the smooth transition animations we use, are demonstrated in our accompanying video.

\section{Conclusion}

We have introduced a new method, \emph{MapTrix}, for visualising many-to-many flows by connecting an OD matrix with origin and destination maps. We have provided a detailed analysis of the design alternatives and have given an algorithm for computing an arrangement with crossing free leader lines.

We conducted two user studies of visual representations of many-to-many flow. 
In our first study we compared MapTrix with a flow map with bundled edges and with OD Maps for different country maps. 
All three visualisations performed well for the smallest data set (AU - 8 locations), but MapTrix and OD Maps were far better for DE and NZ (16 locations). There was no statistically significant difference between MapTrix and OD Maps on data sets of this size.  Surprisingly, we did not find that country shape  affected performance: in particular we had expected this to affect OD Maps.

In our second study we compared MapTrix and OD Maps on two larger data sets (CN - 34 locations and US - 51 locations). 
Both performed relatively well for all tasks and we did not find that one method outperformed the other even for individual tasks. We did find in the pilot that analysing flow between or within regions for data sets of this size was extremely difficult with both methods, though slightly easier with OD Maps. Thus, in the study we used highlighting to help with analysis of regional flow.

In the first study users ranked MapTrix highest in terms of design and readability while in the second study  MapTrix is preferred for design but OD Maps for readability. 

The designs presented in this paper and our user study concentrate on static visual representations of dense many-to-many flows. However in our second user study we did explore the usefulness of highlighting for analysis of regional flow. 
The results of our studies led us to implement several types of interaction, not only highlighting but also filtering and region zooming.  
We plan to evaluate these in our future work. One limitation of our studies is that participants were predominately students or researchers: we plan further evaluations with domain experts.

\acknowledgments{
  Data61, CSIRO (formerly NICTA) is funded by the Australian Government through the Department of Communications and the Australian Research Council through the ICT Centre for Excellence Program. We would like to thank Dr Aidan Slingsby and other members of the giCentre, City University London for their work and discussions on OD Maps. We thank Dr Haohui (Caron) Chen and all of our user study participants for their time and feedback.
}

\bibliographystyle{abbrv}
\bibliography{template}
\end{document}